\PassOptionsToPackage{table}{xcolor}
\documentclass[]{svjour3} %EMSE
\usepackage[utf8]{inputenc}
\usepackage{todonotes}

\RequirePackage[abbreviate=true, style=numeric,dateabbrev=true, isbn=false, doi=true, urldate=comp, url=false, maxbibnames=9, backref=false, backend=biber]{biblatex} %ACM-Reference-Format
\usepackage{booktabs}
\usepackage[space]{grffile} % space in fig names
\usepackage{graphicx}
\usepackage{hyperref}
\usepackage{appendix}

\newcommand{\bi}{\begin{itemize}}
\newcommand{\ei}{\end{itemize}}
\newboolean{showcomments}
\setboolean{showcomments}{true} % toggle to show or hide comments
\ifthenelse{\boolean{showcomments}}
    {
        \newcommand{\jeff}[1]{\textcolor{cyan}{{\it [Jeff says: #1]}}}
        \newcommand{\daniel}[1]{\textcolor{blue}{{\it [Daniel says: #1]}}}
        \newcommand{\marco}[1]{\textcolor{red}{{\it [Marco says: #1]}}}
         \newcommand{\neil}[1]{\textcolor{red}{{\it [Neil says: #1]}}}

    }
    {
        \newcommand{\jeff}[1]{}
        \newcommand{\daniel}[1]{}
        \newcommand{\marco}[1]{}
        \newcommand{\neil}[1]{}
    }
\newcommand{\replPackage}[0]{\url{http://doi.org/10.5281/zenodo.4568517}}
\journalname{Empirical Software Engineering}

\title{Understanding Peer Review of Software Engineering Papers}
\author{Neil A. Ernst \and Jeffrey C. Carver \and Daniel Mendez \and Marco Torchiano}
\date{August 2020, revised February 2021}
\institute{University of Victoria, Canada \and University of Alabama, USA \and Blekinge Institute of Technology, Sweden, and fortiss GmbH, Germany \and Politecnico di Torino, Italy}

\begin{document}

\maketitle

\begin{abstract}
Peer review is a key activity intended to preserve the quality and integrity of scientific publications. 
However, in practice it is far from perfect.

We aim at understanding how reviewers, including those who have won awards for reviewing, perform their reviews of software engineering papers to identify both what makes a good reviewing approach and what makes a good paper.

We first conducted a series of interviews with recognised reviewers in the software engineering field. 
Then, we used the results of those interviews to develop a questionnaire used in an online survey and sent out to reviewers from well-respected venues covering a number of software engineering disciplines, some of whom had won awards for their reviewing efforts.

We analyzed the responses from the interviews and from 175 reviewers who completed the online survey (including both reviewers who had won awards and those who had not). We report on several descriptive results, including:
Nearly half of award-winners (45\%) are reviewing 20+ conference papers a year, while 28\% of non-award winners conduct that many. The majority of reviewers (88\%) are taking more than two hours on journal reviews. We also report on qualitative results. 
Our findings suggest that the most important criteria of a good review is that it should be factual and helpful, which ranked above others such as being detailed or kind. 
The most important features of papers that result in positive reviews are a clear and supported validation, an interesting problem, and novelty. 
Conversely, negative reviews tend to result from papers that have a mismatch between the method and the claims and from papers with overly grandiose claims. Further insights include, if not limited to, that reviewers view data availability and its consistency as being important or that authors need to make their contribution of the work very clear in their paper. 

Based on the insights we gained through our study, we conclude our work by compiling a proto-guideline for reviewing. One hope we associate with our work is to contribute to the ongoing debate and contemporary effort to further improve our peer review models in the future.

\keywords{peer review,  interview, survey}
\end{abstract}

\section{Introduction}
Peer review is a critical aspect of scientific practice. Reviewers are an important set of gate-keepers for the quality and integrity of publications. While editors and PC chairs in the software engineering research community 
%\footnote{In the following text, unless we specify exactly, we use the word `decision-maker' to refer to either PC chairs or journal editors/associate editors.} 
are ultimately responsible for publication decisions, it is the reviewers who spend the most time with a submitted manuscript and who deliver editorial recommendations to the authors and decision-makers. The roles and responsibilities of the reviewers may differ according to the venue, the editorial team, and the reviewing regime, but typically the reviewer feedback is advisory and helps the editor make an informed decision whether to accept a submitted manuscript.

A good reviewer also improves the quality of scientific practice. Ideally, the authors of a manuscript are able to improve it based upon the review feedback, be it in the form of better presentation, additional related work, different methodological approaches, or by strengthening data sources. There are numerous perspectives on the proper role for peer review and what peer review ought to emphasize. We provide an overview of these perspectives in Section~\ref{sec:related}. Typically, for technical scientific results, peer reviewers are often expected to comment on the novelty of the work, the soundness and reliability of the work, and the presentation of the work. These expectations are summarized by the maxim ``is it new, and is it true?"~\cite{macauley12}. 

Even with the importance of peer review described above, it is notoriously imperfect. The NeurIPS experiment~\cite{nips14} assigned the same set of submitted papers to two separate review committees of similar composition. The experiment then analyzed the degree of consistency in the recommendations made by each committee (i.e., would the committee accept or reject a given paper). In a perfect world, we would hope for 100\% consistency, that is, the two independent review committees arrive at the same decision on each paper. In practice, however, the two committees disagreed on the decision for 26\% of the papers, i.e., committee 1 decided to reject when committee 2 decided to accept, and vice versa. Of the papers ultimately accepted, the two PCs disagreed on 57\%. This result implies a substantial degree of randomness to the process which contravenes the stated goals of peer review. 

To better understand the work of individual peer reviewers in software engineering, what approaches they use, and their underlying worldviews and assumptions, we conducted an exploratory study. In this study, we sought to understand some of the existing issues with peer review in software engineering, and ideas for improvement. In particular, we considered three research questions:

\begin{description} % in reviewer's opinion removed
    \item[\textbf{RQ~1}] What practices do reviewers follow when deciding whether to accept an invitation, conducting the review itself, and writing up the results of the review? %14-18,24,33,36
    \item[\textbf{RQ~2}] What are best practices in reviewing with respect to process, method, and content? %. 25,31, qual: Q35 and 53
    %%MTk: removed "In their opinion," since they are all opinions 
    \item[\textbf{RQ~3}] What characteristics of a paper lead to a more favourable review?
    % a paper more likely to be accepted (or lead to a positive revieware best practices in paper writing? % 19,22,23, qual: 34 and partly Q22, Q23
\end{description}

To answer these research questions, we first conducted a series of interviews with internationally recognized reviewers in the software engineering field. 
Then, we used the results of those interviews to develop a questionnaire for an online survey. 
We sent out our questionnaire to reviewers from well-respected venues covering a number of software engineering disciplines, some of whom had won awards for their reviewing efforts. Section~\ref{sec:ResearchDesign} provides more details about our approach. 

In this manuscript, we report on the results and discuss their implications. Our paper's main contributions are 1) a detailed analysis of the way in which peer reviewers in software engineering operate; 2) two concrete guidelines for peer review in software engineering, based on the responses of these reviewers. One is for program chairs and editors, and the second is for reviewers in software engineering. We additionally disclose our (anonymized) data set as well as other artifacts to the research community to allow for further exploratory analyses. The replication package can be found at \replPackage{}.

One hope we associate with our work is to contribute to the ongoing debate on how to further improve our peer review models in the future.

\section{Research Design: Multi-Study Approach}
\label{sec:ResearchDesign}
We designed our study as a multi-study in which we began with a set of interviews of selected reviewers followed by an informal analysis of the interview results. We used this analysis to generate a more detailed questionnaire for use in a broad online survey of reviewers. We then elaborate on those survey results using coding and statistical and descriptive analyses. The remainder of this section provides more details on our approach.

\subsection{Interview Study}

We began our study by conducting a series of interviews with software engineering researchers who had won at least one award for their review practices at ICSE, MSR, ICSA, ICSME, or SANER. We contacted these individuals opportunistically, using personal contacts (i.e., convenience sampling). During the first contact, we provided the participants with an overview of the goals of our study and the topics we would be covering in the interviews. Table~\ref{tbl:interviewees} lists the demographics of our interviewees (while preserving their anonymity).

We developed our questionnaire in a curiosity-driven manner, by jointly compiling a broad spectrum of questions covering the following general areas:
\begin{itemize}
    \item How the interviewees were introduced into reviewing, general (topic) preferences, and reviewing practices.
    \item Experiences and opinions on what constitutes a good paper and what  constitutes a good review.
\end{itemize}
The resulting questions we used as orientation are available in the replication package. The interviews were semi-structured to allow for deviations from the originally planned questions. 

Each author then conducted two interviews. We conducted all but one interview either via phone (P1 and P6) or via a video conferencing tool (P3, P4, P5, P7, and P8), with the remaining interview done by email (P2). The interviews ranged from 45-60 minutes. All interviews followed the script available in our online material (\replPackage{}). On average, our interviewees review 64 papers a year, taking approximately 150 minutes (2.5 hrs) per review, and they all share that they have been gradually introduced into peer reviewing by their advisors as PhD students by co-reviewing manuscripts.  

\begin{table*}
    \centering
        \caption{Interview Participant Demographics. RpY: reviews per year; Start: year begun as reviewer; TpR: time per review in minutes. Note that we removed who won which award to support the anonymity of our interviewees. 
        } 
        \label{tbl:interviewees}
            \rowcolors{2}{gray!25}{white}
    \begin{tabular}{ccccp{3cm}p{3.5cm}}%}
        \toprule
%        \rowcolor{gray!50}
    & RpY & Start  & TpR & Expertise & Reviewing medium   \\
    \midrule
    P1 & 50-70 & 2009  & 60-120 & repository mining,  empirical SE & Print out  \\
    P2 & 60 & 2012  & 180 & repository mining, empirical SE & n/a  \\
    P3 & 100-120 & 2003  & 120 & empirical studies, mining, and software evolution & iPad+pencil  \\
    P4 & 60 & 2006 &  180-240 & architecture, design & Annotated pdf for conference, print out for journal  \\
    P5 & 100 & 1996  & 180 & testing, empirical SE & Ipad/annotated pdf  \\
    P6 & 30 & 2006 &  120 & code analysis, scaling to practice & Print out  \\
    P7 & 30-40 & 2002 & 60-120 & human studies, qualitative & Print out, let it sit 1 day. Read all papers before reviewing \\
    P8 & 57 & 2001 & 180-240 & empirical, surveys & Text editor + PDF \\
    \bottomrule
\end{tabular}
\end{table*}

To analyse the data, we intentionally did not follow a specific coding approach.
Because the interviews were semi-structured, only portions of the data are comparable. Our idea was to use the general insights from the interviews to steer the development of our survey questionnaire. Therefore, each author summarised their key insights from their interviews and we discussed those insights and our general impressions in joint meetings. After the eighth interview, we were confident we had enough information to design the online survey.

\subsection{Online Survey}

Following the interviews and informal analysis, we created a Qualtrics survey following guidelines from Kitchenham and Pfleeger~\cite{kitchenham2002principles} and Smith et al.~\cite{Smith:2013wp}. % Graziotin  July 17-Aug 8
% Alexey 
% Courtney
% can't find the others! 

% finalize survey Jun 2018
% first email Oct 4 2018
% reminder Oct 15?

We then piloted a survey instrument with 3 respondents. The first 2 respondents helped us refine the instrument (pre-test) and the last respondent validated how well the instrument answered our research questions. The final survey instrument appears in the replication package. We solicited contact information for reviewers from top software engineering venues including the 2017 and 2018 editions of EASE, ESEM, ICSE, FSE, and the editorial boards of JSS, TSE, TOSEM, and EMSE. 

We contacted a total of 985 reviewers and sent one reminder email in addition to the initial recruitment email (detailed in the replication package). This sampling frame represents a reasonable proportion of the active reviewers in our field. We do not know the total number of unique reviewers in the software engineering community. However, if there were approximately 9000 software engineering \emph{authors} in 2012~\cite{Fernandes2014}, our sampling frame is likely close to 10\% of authors, and of those authors, a substantial number will not be reviewers (e.g., junior authors or one-off papers).

In the remainder of the paper, we structure the results around the research questions. Most of our analysis relies on descriptive statistics and manual coding of open-ended questions (where respondents can answer questions in a textual manner). We performed the coding without a reference model and assigned codes to individual categories followed by a simple categorisation. Two coders individually coded the responses and cross-validated each other's work. 
A different person conducted spot checks of the coding. This check was done by going through a sample of codes and the underlying data (covering approximately 10\% of the data), followed by a joint discussion to clarify potential doubts. Our replication package includes the survey questions, the raw data from the survey, and the coding results. Finally, the institutional ethical approval certificate by the first two authors is available as part of the replication package. 

\section{Research Questions and Results}
In this section, we report on the demographics of the respondents followed by the results for each of our three research questions. We discuss the implications of those results immediately after each research question. We then highlight a few other results that reflect emerging topics in peer review in software engineering.
 
\subsection{Demographics} % survey 1-10
\label{sec:demo}

Of the 985 invitees, 229 clicked on the Qualtrics link (23.2\%). We recorded 201 unique responses (by IP), as well as 2 emailed responses. Before proceeding with any analysis, we checked the completeness of the responses. We consider a response complete, and kept it in the analysis, if the respondent answered at least $2/3$ of the questions. Using this threshold, we removed 26 responses, keeping 175 for the full analysis. Here, we report some demographics on the respondents.

%out of which we considered 175 responses to include sufficient information to be included in our analysis (we excluded responses that were mostly incomplete or contained noisy answers). 

% \begin{figure}[htb]
%     \centering
%     \includegraphics[width=.6\linewidth]{figures/completeness-1.pdf}
%     \caption{Completeness of answers}
%     \label{fig:completeness}
% \end{figure}

First, regarding recognition of the participating reviewers, 35\% of the respondents had received some form of \emph{best reviewer award} (19\% for conference reviews, 12\% for journal reviews, and 3\% for both). Best reviewer awards, common in most software engineering research conferences, typically reward 5-10 reviewers who best helped in the review process (usually with lengthy, detailed, timely reviews). In addition, 73\% of the respondents had served as either a conference Program Chair or a journal Editor. These respondents were twice as likely to have received a best reviewer award as the other respondents.

%unnecessary
%\begin{figure}[htb]
%    \centering
%    \includegraphics[width=.5\linewidth]{figures/treemap-award-1.pdf}
%    \caption{Award recipients}
%    \label{fig:awards}
%\end{figure}

\begin{figure}[htb]
    \centering
    \includegraphics[width=.8\linewidth]{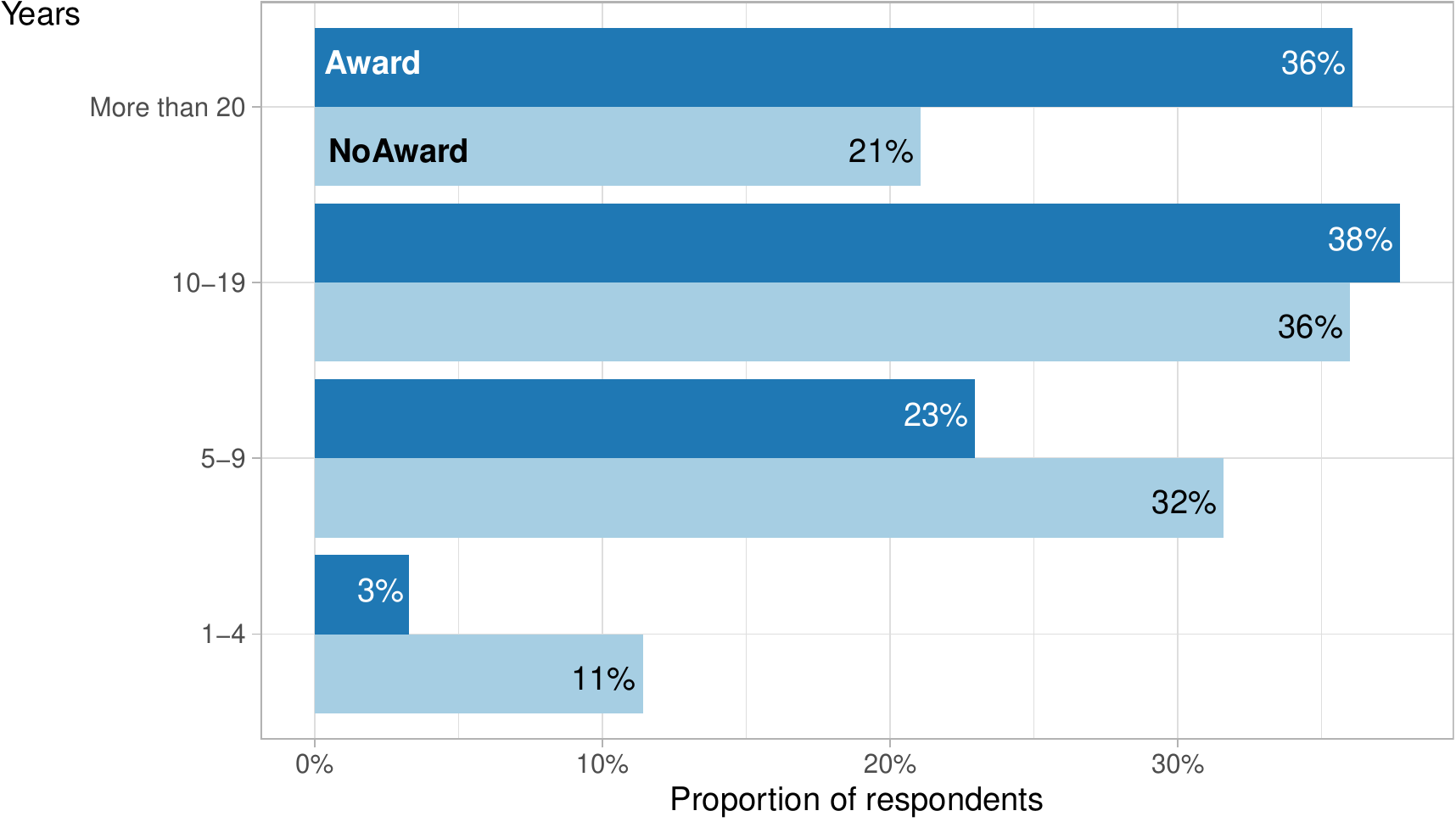}
    \caption{Reported experience in reviewing papers.}
    \label{fig:experience}
\end{figure}

Second, regarding years of experience, the median experience in reviewing papers is between 10 and 19 years. Figure~\ref{fig:experience} shows a large difference between award recipients and the others, with 36\% of recipients reporting more than 20 years of experience compared with 21\% of non-recipients and only 3\% of the award recipients reported less than 4 years of experience compared with 11\% of the non-recipients.

\begin{figure}[htb]
    \centering
    \includegraphics[width=.99\linewidth]{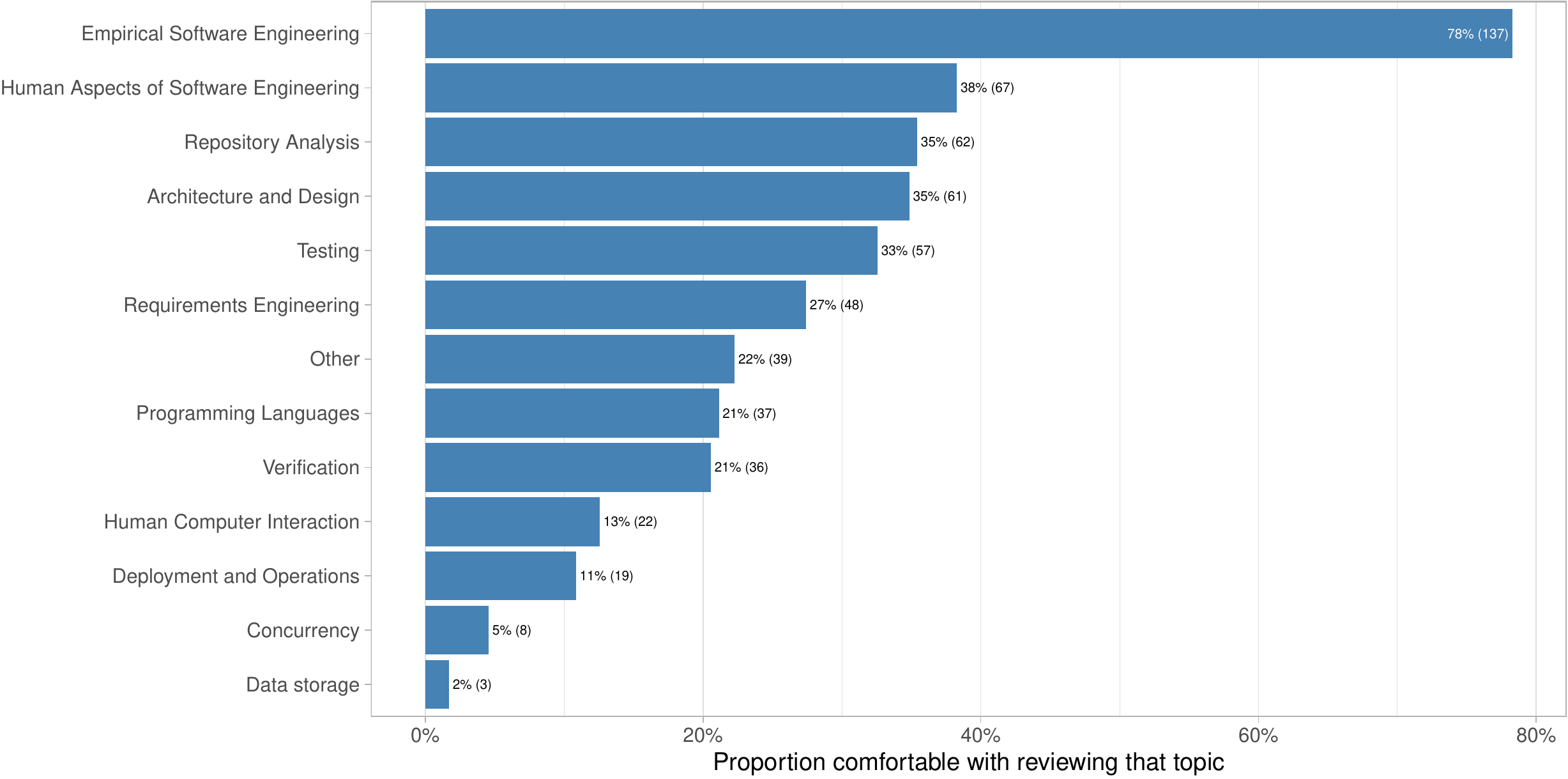}
    \caption{Topics respondents are comfortable reviewing.}
    \label{fig:topics}
\end{figure}

Third, regarding knowledge of software engineering topics, Figure~\ref{fig:topics} shows that the respondents are knowledgeable about several aspects of software engineering, with empirical SE, human aspects, and repository analysis being the most common. This result reflects our sampled venues such as EASE, TSE, and ICSE.

\begin{figure}[htb]
    \centering
    \includegraphics[width=.9\linewidth]{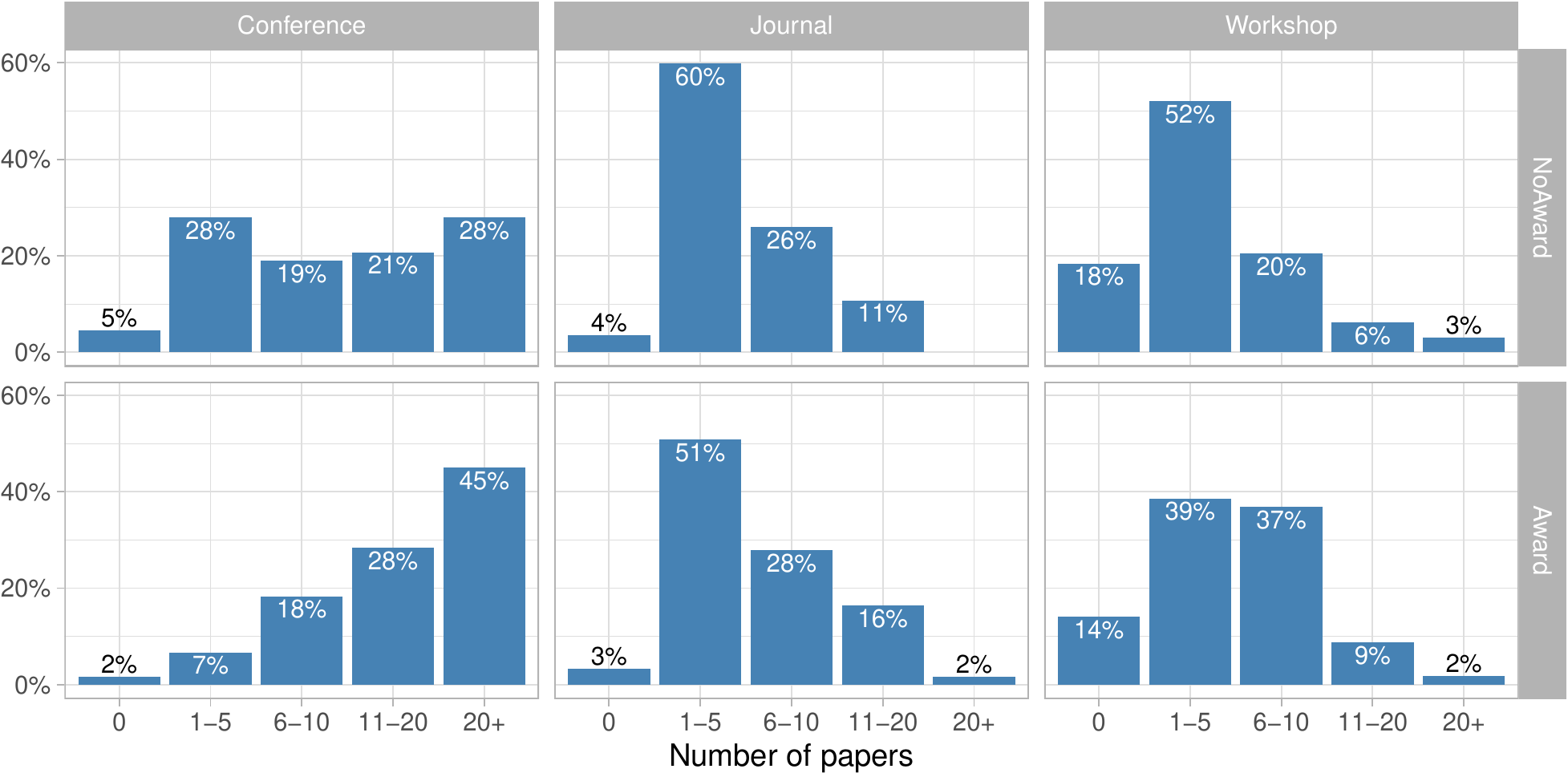}
    \caption{Review load distribution for different venues. Rows reflect non-award (top) and award winning respondents.}
    \label{fig:load}
\end{figure}

Finally, in terms of reviewing load, Figure~\ref{fig:load} shows that award winners tend to do more conference reviewing (45\% review more than 20 papers a year) and more reviewing in general. This result may also be confounded with reviewer experience.

\subsection{RQ1: Reviewer Practices} %14-18,24,33,36
Our first research question, \textit{What practices do reviewers follow when deciding whether to accept an invitation, conducting the review itself, and writing up the results of the review?} examines the mechanics of the review process. In this section, we divide these results into three parts, according to the review process: review invitation, conducting the review, and writing the review.

\paragraph{Review invitation} 
%- Reviews per year %10
As noted in the demographics, the respondents are very active reviewers. For journals, 39\% conducted more than 6 reviews per year. For conferences, 57\% conducted 11 reviews per year, with 34\% doing more than 20 reviews per year. 

%- Guidelines %24

\begin{figure}[htb]
    \centering
    \includegraphics[width=.95\linewidth]{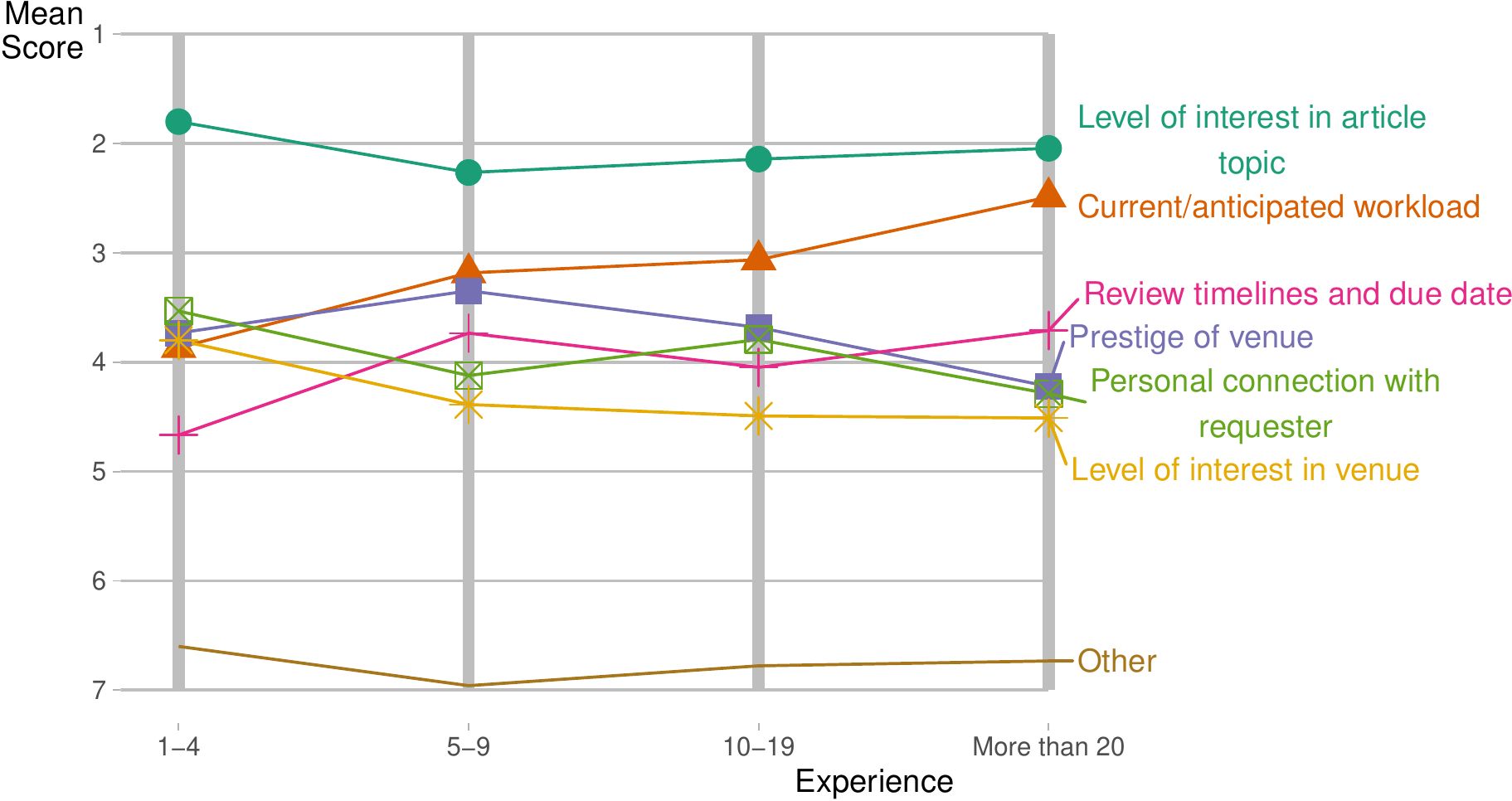}
    \caption{Reasons for accepting journal review requests, by reviewer experience. Scores reflect mean rank by survey respondents, 1=most important.}
    \label{fig:journal-accept}
\end{figure}

%- Invites %11-13
To understand these heavy reviewing loads, we asked our respondents how they decide when to accept an invitation to review a paper, join a PC, or bid on a paper. The commonly accepted practice is to review as much as you are reviewed. For a typical conference paper submission that receives three reviewers, one would expect to perform three reviews in return. Thus, a review load of 20 or more conference reviews implies that reviewer may submit 7 or more papers per year. Figure~\ref{fig:journal-accept} illustrates the reasons why reviewers accept journal invites. The most important factors are: their level of interest in the topic and then their current or anticipated workload. Similar results hold for conference PC invitations. Unsurprisingly, as reviewers gain experience, the workload impact becomes more important, and prestige less so. 

\begin{figure}[htb]
    \centering
    \includegraphics[width=0.8\linewidth]{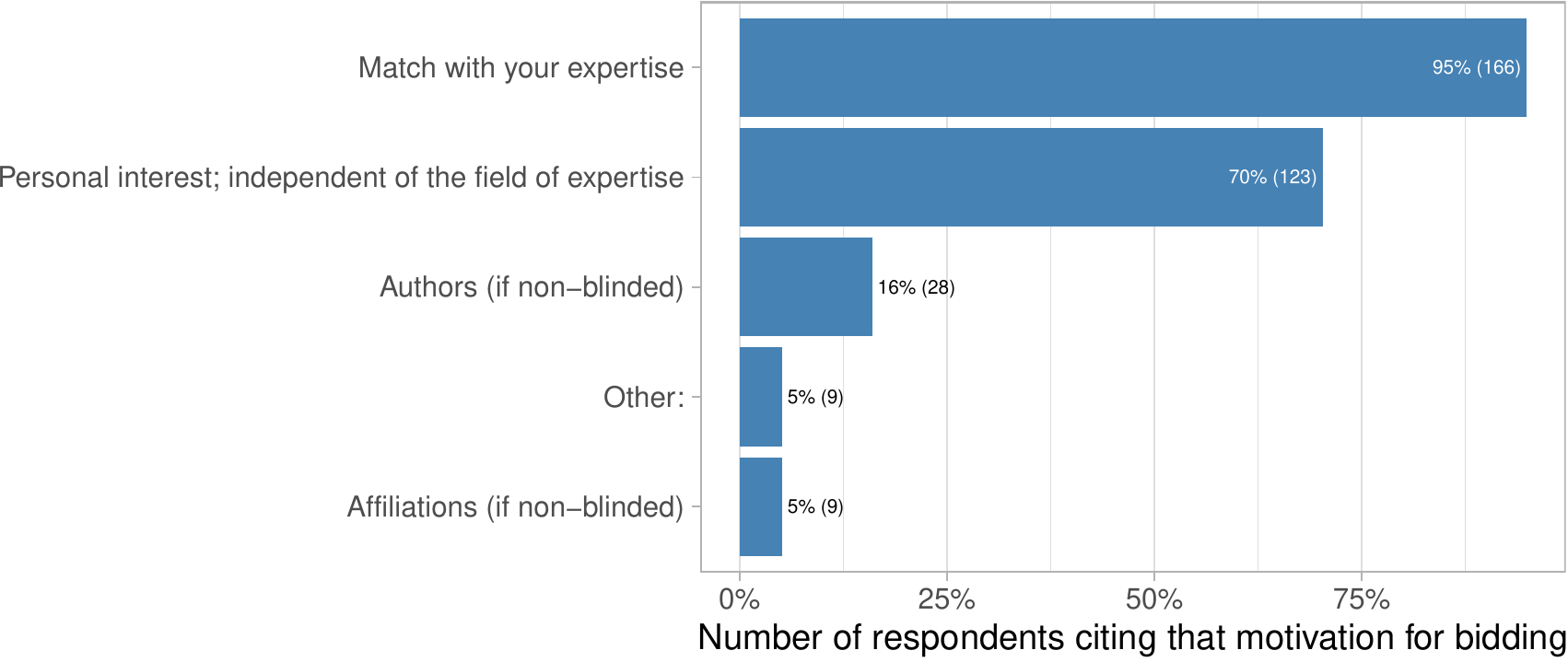}
    \caption{Motivation for bidding on conference papers.}
    \label{fig:pc-bid}
\end{figure}

Figure~\ref{fig:pc-bid} illustrates the reasons reviewers give for bidding on papers as part of a conference review process. These bids guide paper assignment by the program chairs. As expected, most respondents look for a match with their expertise. Personal interest also factors quite high. %, which, while realistic, is not necessarily useful to the PC chairs. 
In some cases, the motivation for bidding is less clear. While it makes sense for a reviewer to bid based on their expertise and ability, as one of the interviewees said, ``\textit{sometimes I will intentionally select papers that I know are not going to get in ... [that I know] this is an easy reject because I want to lighten my reviewing load.}$^\mathsf{P1}$".

%Figure~\ref{fig:pc-bid} shows, indeed, that

\begin{figure}[htb]
    \centering
    \includegraphics[width=0.9\linewidth]{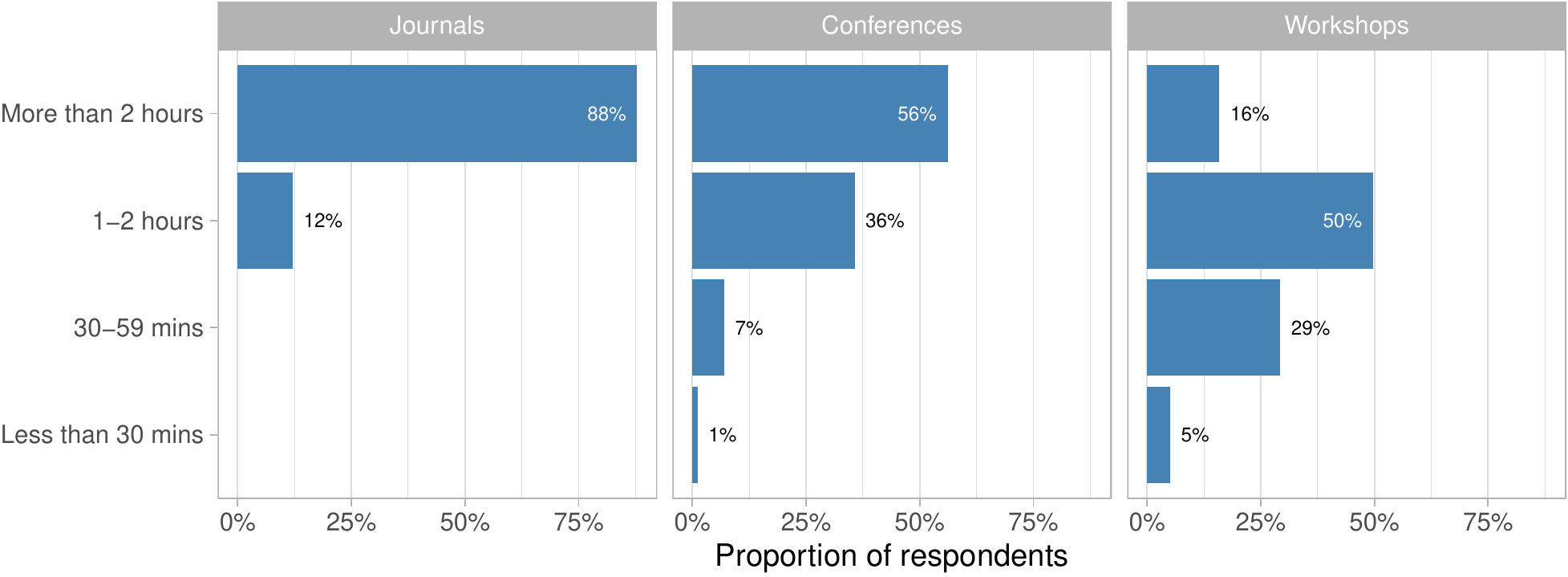}
    \caption{Time spent on reviews, by venue type.}
    \label{fig:time-spent}
\end{figure}

\paragraph{Conducting the review}
Concerning the mechanics of reading and reviewing the paper, we examine time, reading approach, and hardware support. First, Figure~\ref{fig:time-spent} shows the breakdown of time spent performing reviews. Journal reviews typically require over 2 hours (including reading paper and writing review), with conference papers overall taking a bit less time. Because the figures are self-reported, we have to be cautious with their accuracy as respondents may not accurately remember the exact time for the review process.

Second, when asked how many times they read a paper during the review process, a large share of our respondents only does so once (63\% for conferences, 48\% for journals). When we examine this result based on reviewer experience, there is a clear trend that reviewers with more experience require fewer reads of a paper. 

\begin{figure}[htb]
    \centering
    \includegraphics[width=0.8\linewidth]{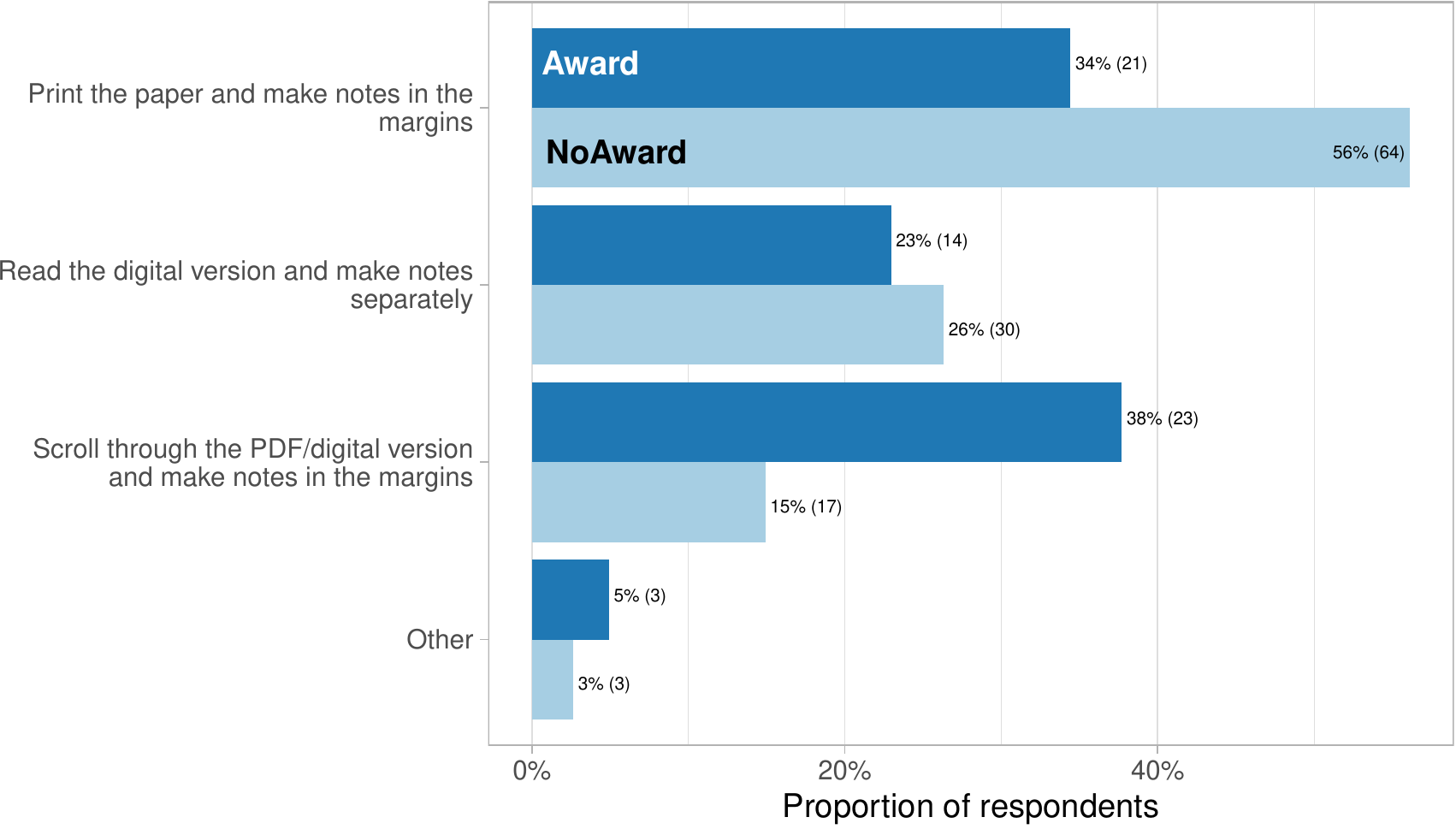}
    \caption{Device support for reviews.}
    \label{fig:devices}
\end{figure}

Finally, Figure~\ref{fig:devices} illustrates the responses to the question about the devices reviewers use to conduct reviews. Interestingly, award winners prefer to use digital approaches (e.g., a PDF reader with note support), although some of our interviewees are pen-and-paper adherents, particularly for reviewing while traveling. Overall, 48\% prefer printed versions. Reviewer experience does not provide any additional insight into this result.

%- Time spent %14
%- Reading order %16
%- Devices and approach %15
\paragraph{Student Involvement in Reviewing}
When it comes to writing the review itself, we examined the role of students in peer review. Many interviewees mentioned their time as a PhD student as the main way they learned how to perform reviews. There is clearly a training benefit to allowing students to participate. At the same time, most conferences explicitly say the reason one is asked to join the program committee is for one's expertise, and consequently, delegating reviews is not allowed. Recently, conferences such as ICSE have begun providing explicit guidelines that allow student feedback, still requiring the invited program committee member to write the review. MSR 2021 used a shadow PC composed of students. 

Our results showed only 24\% of respondents did not use students in their reviews. A larger share (34\%), however, did allow students to review the manuscript separately, prior to meeting to reconcile the student's review and their review. This approach seems to provide a training opportunity. As one interviewee said:
``\textit{my PhD advisor asked [me] to review papers, then went through reviews with [me]...}$^\mathsf{P8}$'' 
% ``\textit{we discuss their review so they learn how to do that and if a point comes to my mind I add it to my own review and add student as the subreviewer}$^\mathsf{P5}$'' 
(some conferences acknowledge sub-reviewers in the proceedings).

%The following section (Section \ref{sec:good-review}) discusses what should actually be in the review. 
%- Students %18
%- Other quality %33
%- Items in review %36

Finally, we asked a Likert-scale question about the importance that certain aspects appear in a submitted peer review. The top four aspects were soundness, points against, significance, and points in favour, with the bottom three being grammatical mistakes, bibliography errors, and typographic mistakes. 

\paragraph{Discussion---Practices} 
We cataloged a set of different practices reviewers followed. One point that stands out is that reviewers blend expertise and interest (independent of expertise) when bidding, which may lead to less expert analysis of papers. 
One explanation for this may be the frequent new topics that emerge in software engineering, such as SE for machine learning systems. There are also clear distinctions between conferences and journal reviews, unique to computer science venues. Conference reviews focus on making decisions in a time-constrained manner.  Journal reviews, on the other hand, are done less frequently, but are longer and take more time.``\textit{[My journal review involves] very profound analysis (even by checking the proofs) with possibility to openly ask questions to authors (rebuttal)''}.$^\mathsf{P5}$

\subsection{RQ2: Best Practices for Conducting Reviews} %. 25,31, qual: Q35 and 53
\label{sec:good-review}

For this research question, we focused on understanding the best practices for \emph{conducting reviews}, with respect to process, method, and content. Within this question, we examine four topics: how reviewers validate associated artifacts, how they assess alignment of the research strategy and the stated problem, the characteristics of an impactful paper, and the characteristics of a good review (both from a reviewer's perspective and from an author's perspective). 

\paragraph{Validation of associated artifacts}
A majority of software engineering research papers are computational data studies (e.g., a study validating new test approaches)~\cite{storey2019methodology}. 
For these papers, as well as for papers conducting survey research and sampling (such as this one!), the authors are often recommended to provide replication packages~\cite{fernndez2019open} to support reproducibility. 
This emphasis has been also a main focus of the ROSE festival (Rewarding Open Science in Software Engineering), which has already appeared at ICSE, FSE, RE, and ICSME\footnote{\url{http://tiny.cc/rosefest}}.
However, these additional artifacts increase the review effort, because they form part of the paper's chain of reasoning and, to some extent, require their own verification during peer review. This additional verification is often done in artifact evaluation tracks of conferences or the open science initiatives of journals. Choosing to participate and have one's artifact evaluated is currently voluntary.

\begin{figure}[htb]
    \centering
    \includegraphics[width=0.9\linewidth]{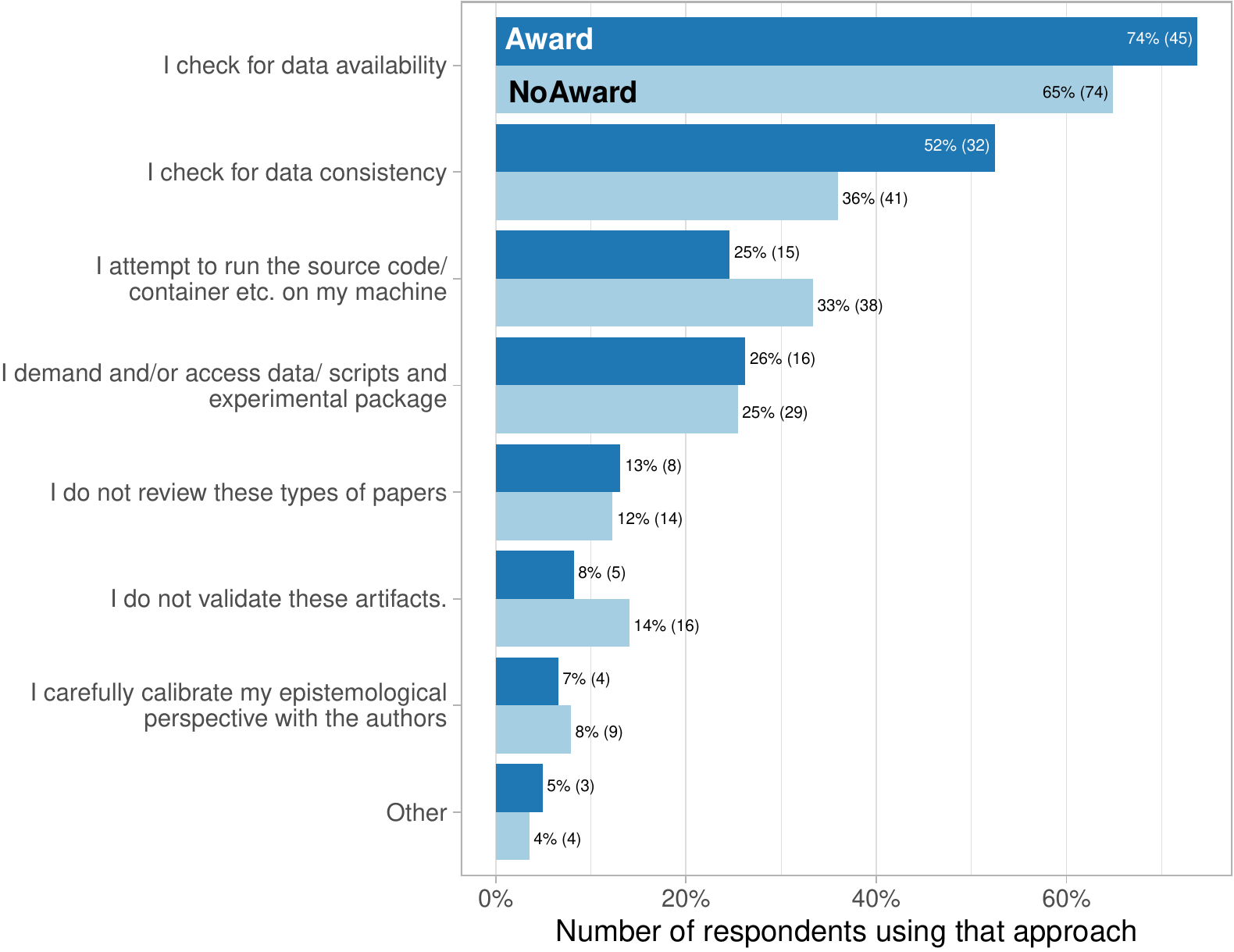}
    \caption{Validate Artifacts and Claims}
    \label{fig:Q25}
\end{figure}

Figure~\ref{fig:Q25} shows the distribution of the approaches respondents used to validate artifacts and claims in papers that include datasets or tools. Overwhelmingly, the respondents performed some type of validation for tools and artifacts, with less than 10\% indicating a total lack of validation.
Most frequently, reviewers seem to check for \textit{availability} and \textit{consistency} when validating artifacts and claims. 
It is less common for reviewers, however, to verify the artifacts themselves. 
Similarly, it is still less common for reviewers to request artifacts that are not provided. 
One of the interviewees summed up the current state as follows:
``\textit{The current way papers are written makes it difficult to reproduce and thus validate the artefacts. But we would not only need to change the authors’ culture, but also the reviewers’ culture.}''

While the overall distribution is similar between those who have won an award and those who have not, there are a few noticeable differences.
Those who have won an award are more likely to check for data availability and consistency.
Conversely, those who have not won an award are more likely to try to run the code locally or to omit the validation completely.

\paragraph{Alignment of research strategy and problem}
We asked our respondents in an open-ended question how they assess the quality of the research method and received a rich picture from 234 responses. % \marco{you mean codes?}. 
Table~\ref{table:method_qual} summarises the codes we attached to at least five responses. 
While the responses included many detailed suggestions, most respondents seem to check whether the choice of research method fits the problem, whether it is in tune with the established literature, whether the level of reported detail allows the reader to understand the research method, and whether the authors clearly describe the threats to validity.

\begin{table}
   \caption{How Reviewers assess the Quality and/or Validity of a Paper's Research Method (showing only selected codes with at least five responses)}
    \label{table:method_qual}
    \rowcolors{2}{gray!25}{white}

    \begin{tabular}{p{.25\textwidth} p{.53\textwidth}r}
    \toprule
     \textbf{Response Code} & \textbf{Explanation} & \textbf{Count} \\  
    \midrule
    Rigor & Assess rigor, validity of methodology relative to published literature or standard practice & 57 \\
     Appropriateness & Is the method appropriate & 27 \\
    Detail & Present appropriate level of detail; Transparent process & 20 \\
    Validity & Discussion of threats to validity & 18 \\
    Conclusions & Do the conclusions follow from the evidence & 10 \\
    Reproducible & Determine if the study is reproducible / replicable & 10 \\
    Support\_material & Includes support materials (e.g. tools, instruments, scripts, ...) & 8 \\
    Knowledge & Use own knowledge to judge validity & 8 \\
    Data\_availability & Availability of data & 5 \\
    Sample & Valid sample & 5 \\
        \bottomrule
    \end{tabular}
   \end{table}
   
One exemplary survey quote reflects well that picture: \emph{``I compare what's in the paper to established guidance for that research method. If the paper deviates from established guidance, I consider whether the deviation makes sense in this context, and whether the authors consciously deviated from the norms for good reasons or just don't know the difference. I look at the chain of evidence from observation to conclusion and consider whether it is clear and unbroken. I consider different quality criteria and threats to validity depending on the type of study and its (often implicit) philosophical position."}

\begin{figure}[!htb]
    \centering
    \includegraphics[width=0.8\linewidth]{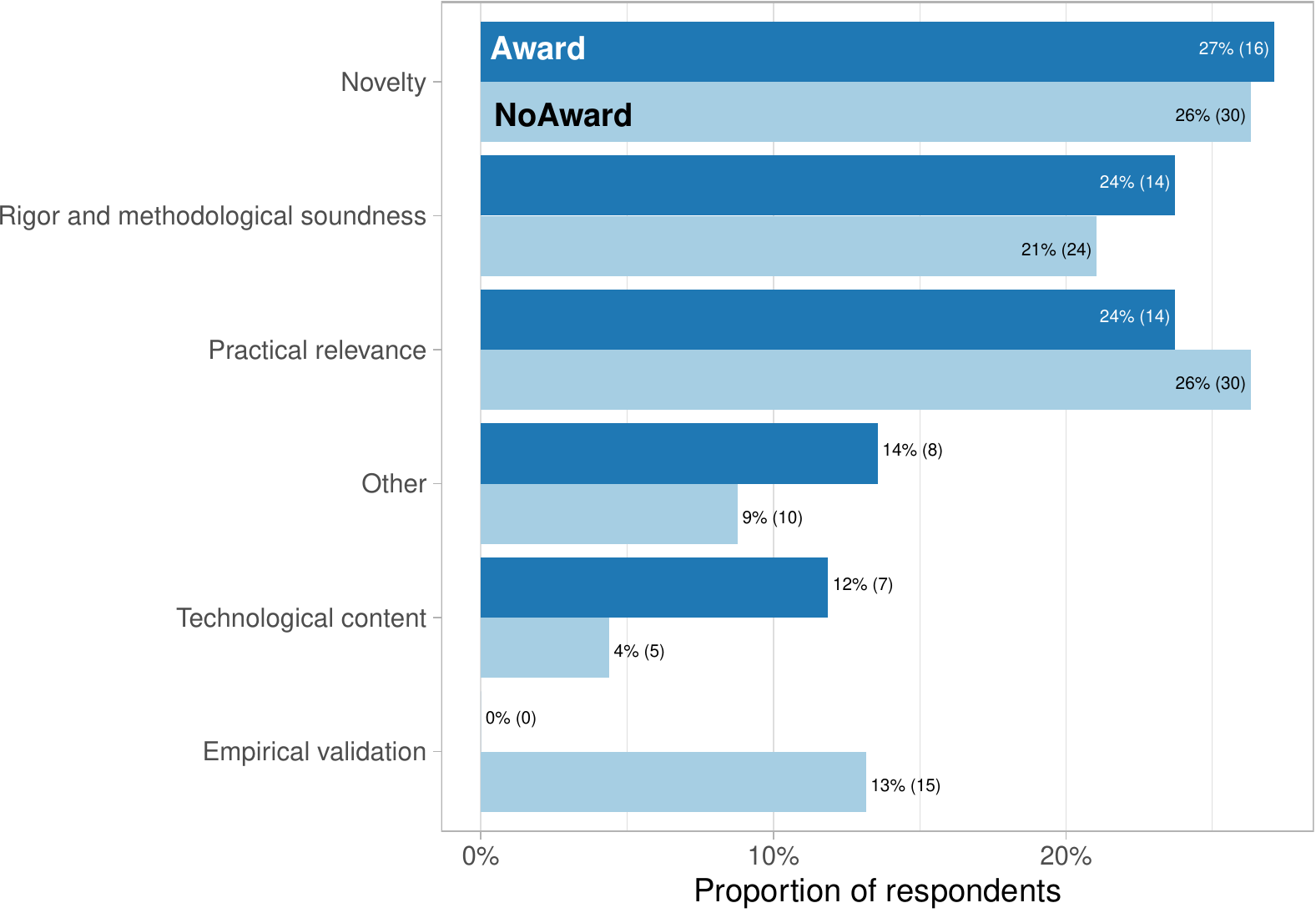}
    \caption{Characteristics of Impactful Papers, Award vs non-Award winning Respondents.}
    \label{fig:Q31}
\end{figure}

\paragraph{Characteristics of impactful papers}
Figure~\ref{fig:Q31} shows the distribution of how respondents judge the impact of papers.
It is interesting to observe that the top three characteristics are \textit{novelty}, \textit{relevance}, and \textit{methodological rigor}, which can sometimes be in conflict with each other. 
There seems to be a balance between the need for relevance and scientific rigor, which has also been a perennial discussion in conferences with industry tracks or sessions.
It is also interesting that 13\% of the respondents who did not receive an award found \textit{empirical validation} to be an important characteristic while none of the respondents who received an award for reviewing thought this characteristics was important.

\paragraph{Characteristics of a good review (Reviewer's perspective)}
We asked respondents to rank the importance of the characteristics of a good peer review.
This question took the form of a drag-and-drop list in our survey tool. Given a randomized list of choices, respondents re-ordered them as they wished, with a choice receiving a score of 1 at the top of the list and 8 at the bottom.

\begin{figure}[!bth]
    \centering
    \includegraphics[width=0.9\linewidth]{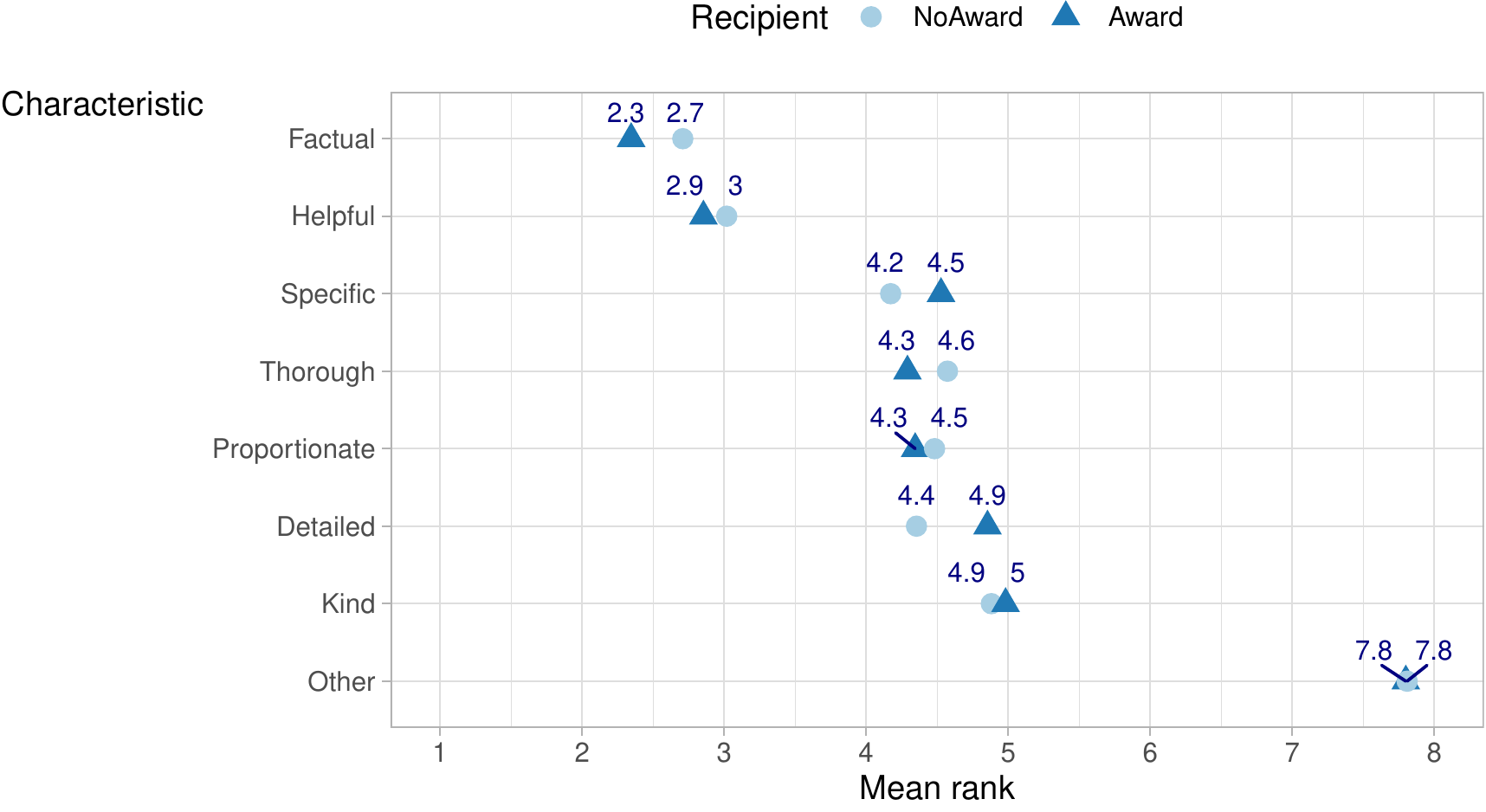}
    \caption{Ranked Characteristics of a Good Peer Review. Ranks closer to the left (1) are better.}
    \label{fig:Q35}
\end{figure}

%Table~\ref{table:Q35} 
Figure~\ref{fig:Q35} shows the results ordered from the \textit{most important characteristic} (lowest number) to the \textit{least important characteristic} (highest number).
While the order differs slightly between the respondents who received an award and those who did not, one consistent result is that \textit{Factual} and \textit{Helpful} are the most common factors.
Interestingly, the gap between \textit{Helpful} and the next answer is much larger than the gap between any other responses.
This result suggests that the top two factors are clearly more important than the others.
A quote from one of the interviewees provides some additional perspective: 
``\textit{I learned two important lessons: 1) no paper is perfect, 2) be always constructive in the reviews, meaning do not only point to issues but explain to the authors how they could deal with each of these issues.}$^\mathsf{P2}$''

%% MTk: replaced with the above diagram
%\begin{table}
%    \caption{Ranked Characteristics of a Good Peer Review. A lower score is better.}
%    \label{table:Q35}
%    \centering
%    \input{table_peer-review}
%\end{table}

\paragraph{Characteristics of a good review (Author Perspective)}
Given that the respondents are also paper authors themselves, we asked them to describe the most important aspects of a high quality peer review, from their perspective as an author.
This information provides an additional perspective into the best reviewing approaches.
Based on the coding of the results, we identified four classes of feedback respondents viewed as important.
Table~\ref{table:Q53} provides a breakdown of the responses for the top two categories of responses.
Across both categories, the factors that are most important focus on reviews being constructive and providing factual, accurate, and fair comments.

The largest category, with 140 responses, related to the \textit{Form and Shape of the Review}.
Within this category, the most common type of answer related to providing \textit{constructive criticism}.
Examples include: \textit{``Concrete, specific suggestions for improvement'', ``clear guidance on how to improve the paper'', and ``actionable feedback''}.
The second most common type of answer in this category related to providing \textit{factual statements}.
Examples include: \textit{``Justified feedback'', ``fairness and factual arguments'', and ``comments backed up with factual evidence''}.

\begin{table}
    \caption{Aspects of High Quality Review}
    \label{table:Q53}
    \rowcolors{2}{gray!25}{white}
\begin{tabular}{p{.25\textwidth} p{.53\textwidth}r}
\toprule
     \textbf{Response Code} & \textbf{Explanation} & \textbf{Count} \\  
     \midrule
     \multicolumn{2}{c}{\textit{Form and Shape of Review}} & 140 \\
     \midrule
    Constructive Criticism & Review should be constructive and give indicators for improvement & 85 \\
    Factual Statements & Statements made by the reviewer should be factual and backed up / justified & 27 \\
    Clarity & The statements and recommendations should be clear and detailed & 13 \\
    Transparency & The decision / recommendation to chair/editor should become evident from the statements and arguments in the review & 4 \\
    Reasonable Suggestions & Suggestions made should be reasonable and feasible & 3 \\
    Comprehensiveness & Review should be in-depth and reproducible rather than being a short, non-helpful statement & 3 \\
    Soundness & Review should be sound and coherent & 3 \\
    Professional Tone & Review should be written in a professional, non-emotional tone & 2 \\
    \midrule
\end{tabular}
\begin{tabular}{p{.25\textwidth} p{.53\textwidth}r}
    \midrule
    \multicolumn{2}{c}{\textit{Conduct and Attitude}} & \hspace{6.3mm}49 \\
     \midrule
    Accuracy & The reviewer took care to read the paper and provide a review that is factually correct relative to the contents of the paper & 14 \\
    Fairness & Reviewer should be fair and carefully consider both negative and positive points & 14 \\
    Expertise & Reviewer should have clear expertise on the topic & 10 \\
    Openness & Reviewer should be objective and open to new ideas that might not fit the reviewer's belief system & 8 \\
    Focus on Soundness & Reviewer should focus on methodological soundness rather than on actual results & 2 \\
    Honesty & Reviewer should be honest about views and expectations & 1 \\
     \bottomrule
\end{tabular}
\end{table}

The second largest category, with 49 responses, related to the \textit{Conduct and Attitude} of the review.
Within this category, the two most common types of answers were \textit{accuracy} and \textit{fairness}.
Examples of the responses we coded as \textit{accuracy} include: \textit{``Evidence that the reviewer has invested significant time trying to understand the paper'', ``Makes me feel the reviewer understood what I did and why and what the point is of it all'', and ``The reviewer reads the paper carefully''}.
Examples of the responses we coded as \textit{fairness} include: \textit{``fairness'', ``it should be specific and fair'', and ``objectivity''}.

The other categories were much smaller: \textit{Overall Process} with 4 responses and \textit{Other} with 2. This result suggests that elements such as timely reviews and helpfulness for editorial decisions were seen as less important (from the author's perspective).
Note that because this question used a free-form response, each response could have produced multiple codes.

\paragraph{Discussion — Review Practices} 
We summarize review practices and elaborate on our recommendations in Section \ref{sec:writingguide}.

\subsection{RQ3: Characteristics Leading To Favourable Reviews}
%\subsection{RQ3: Best Practices for Writing Papers} % 19,22,23, qual: 34 and partly Q22, Q23 (check google drive - provide summary in analogy to the one for RQ2)
We were interested in knowing how reviewers describe a particularly good paper. To this end, we asked four questions:
\begin{enumerate}
    \item Describe how you assess a high quality paper.
    \item Select one to three characteristics of papers that lead to more negative reviews from you.
    \item Select one to three characteristics of papers that lead to more positive reviews from you.
    \item If you could make one recommendation to authors to increase acceptance chances, what would it be?
\end{enumerate}

Question 1 was a free-response question.
For questions 2-4, we offered a multiple-choice list of options followed by a free-text answer option (other).

\paragraph{Characteristics of high quality papers}

Table~\ref{table:Q20} shows the most common results from our coding of the free-text responses to this question.
Overall, the most common characteristics reviewers want in a high quality paper relate to aspects of the quality of the research methodology and how well it is presented in the paper. It is also interesting to note which characteristics were not among the most commonly given answers, including the availability of the data (5), the findings themselves (1), and the generalizability of the results (1).

\begin{table}
    \caption{How Reviewers Assess a High Quality Paper (top 10 responses)}
    \label{table:Q20}
    \rowcolors{2}{gray!25}{white}
\begin{tabular}{p{.25\textwidth} p{.53\textwidth}r}
\toprule
     \textbf{Response Code} & \textbf{Explanation} & \textbf{Count} \\  
     \midrule
    Rigor & Assess rigor, validity of methodology relative to published literature or standard practice & 59 \\
    Appropriateness & Is the method appropriate & 27 \\
    Detail & Present appropriate level of detail; Transparent process & 20 \\
    Validity & Discussion of threats to validity & 19 \\
    Other & & 19 \\
    Reproducible & Determine if the study is reproducible / replicable & 12 \\
    Conclusions & Do the conclusions follow from the evidence & 10 \\
    Support Material & Includes support materials (e.g. tools, instruments, scripts, ...) & 9 \\
    Knowledge & Use own knowledge to judge validity & 8 \\
    Proof & Validity of proof/arguments in a formal method & 6 \\
\bottomrule
\end{tabular}
\end{table}

\paragraph{Characteristics leading to negative reviews.}

\begin{figure}[htb]
    \centering
\includegraphics[width=0.9\linewidth]{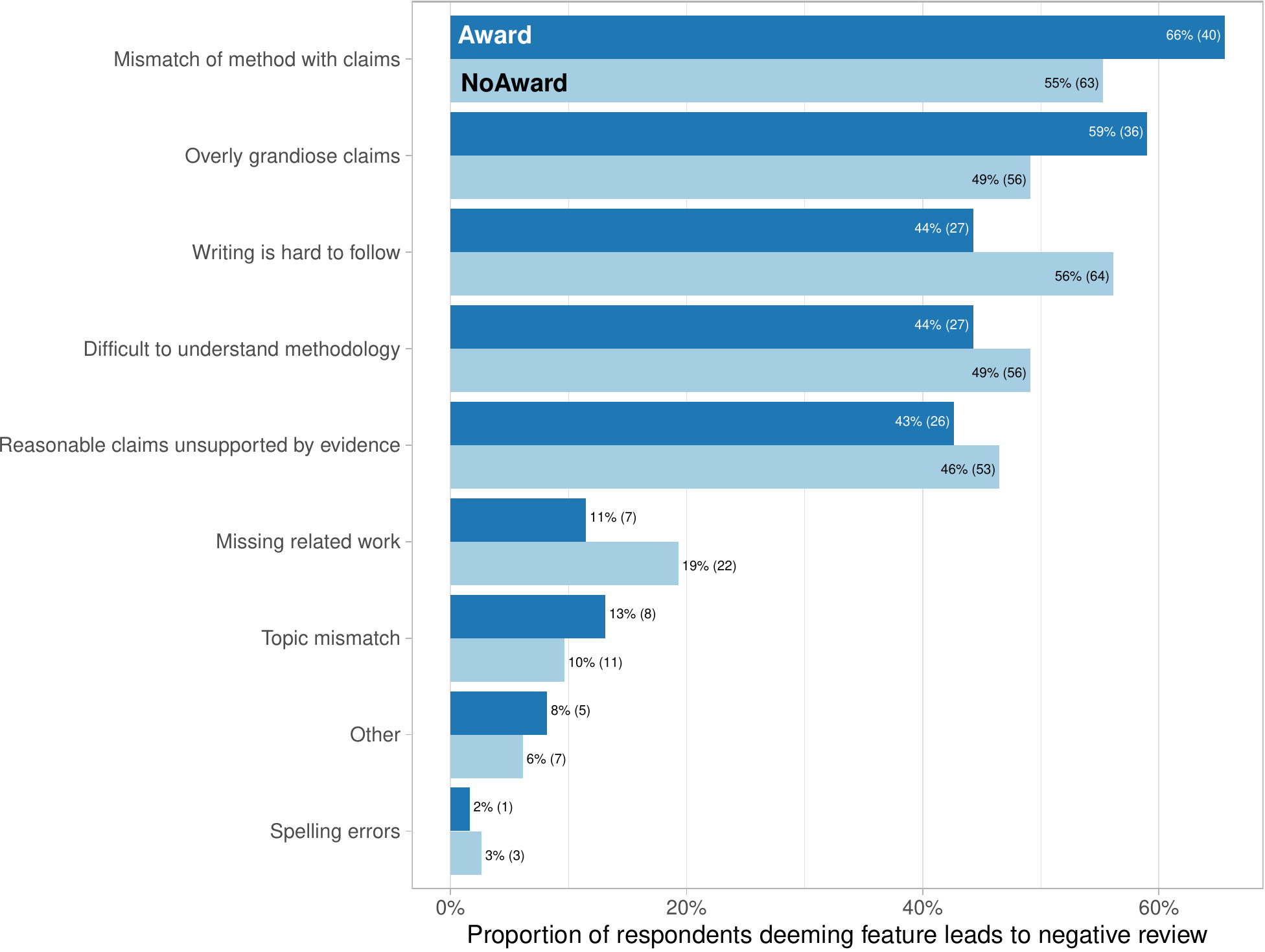}
    \caption{Characteristics leading to particularly negative reviews.}
    \label{fig:Char-negative-reviews}
\end{figure}

Figure~\ref{fig:Char-negative-reviews} shows the ranking of the characteristics that tend to lead to negative reviews.
The highest ranked characteristics include (1) a mismatch of methods with claims raised in the contribution, (2) overly grandiose claims, (3) a writing style that is hard to follow, (4) a research methodology which is difficult to understand, and (5) claims which remain unsupported by evidence. Overall, the differences between those characteristics ranked as important by award winners and the other respondents are rather small.
%However, the following characteristics resulted in slightly larger differences:
%\begin{itemize}
%    \item \textit{Mismatch of method with claims} and \textit{overly grandiose claims} are more important to reviewers who have received an award
%    \item \textit{Writing is hard to follow}, \textit{difficult to understand methodology}, and \textit{unsupported claims} are more important to reviewers who have not received an award.
%    \item Although not ranked as highly important overall, \textit{missing related work} is more important to reviewers who have not received an award.
%\end{itemize}

\paragraph{Characteristics leading to positive reviews.}

\begin{figure}[htb]
    \centering
    \includegraphics[width=0.9\linewidth]{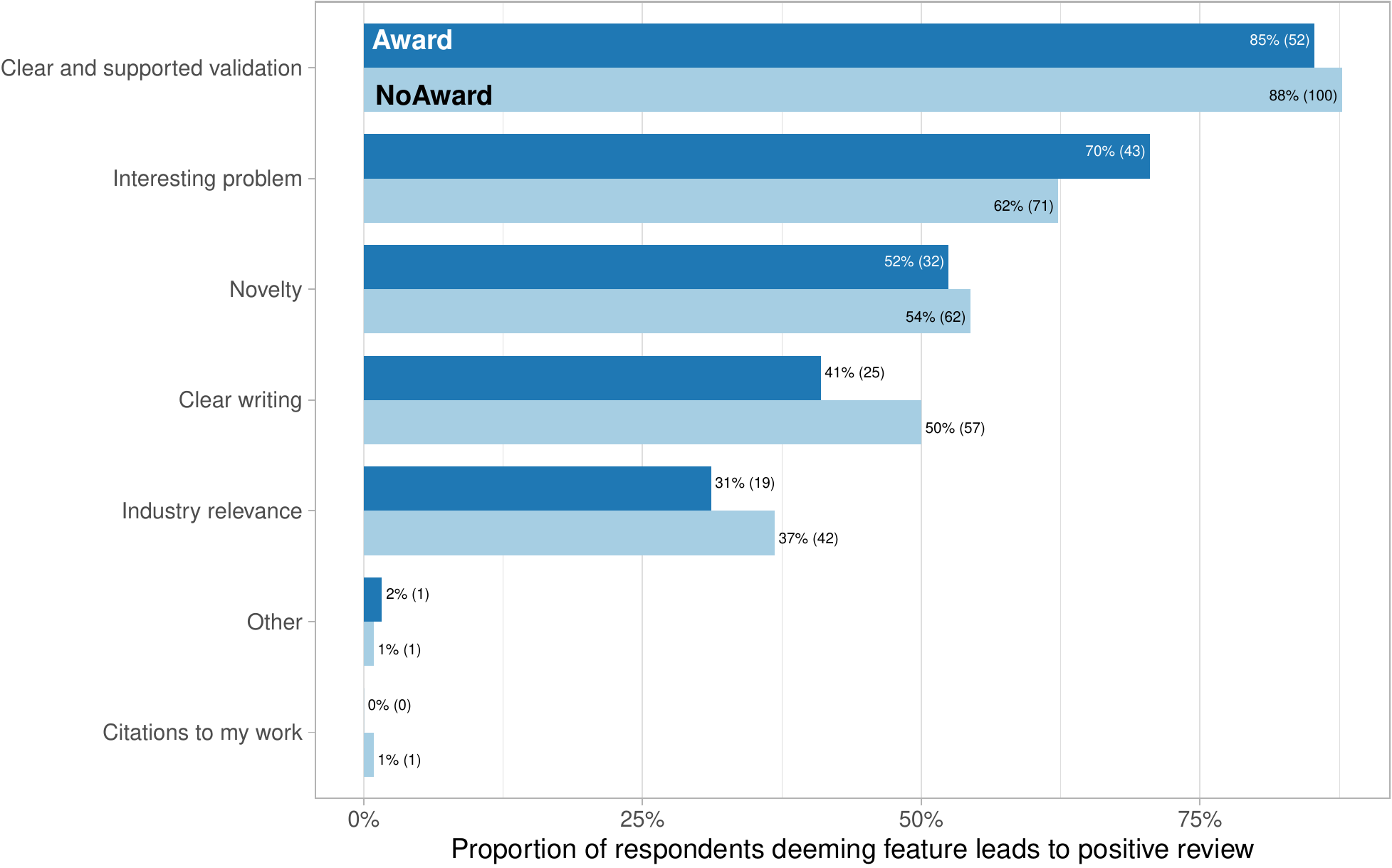}
    \caption{Characteristics leading to particularly positive reviews.}
    \label{fig:Char-positive-reviews}
\end{figure}

Figure~\ref{fig:Char-positive-reviews} summarizes the ranking of the characteristics that tend to lead to positive reviews, using reviewer experience as the independent variable.
In this case, the results are a bit more clear.
The highest ranked characteristic is to offer a clear and supported validation followed by solving an interesting problem, novelty, a clear writing style, and having a high practical relevance. 
%Differences in the ranking are particularly evident in the characteristics clear writing style (favoured by reviewers without an award) and that the problem should be interesting (favoured by those with an award).

\paragraph{Recommendations to authors.}

\begin{figure}[htb]
    \centering
    \includegraphics[width=0.9\linewidth]{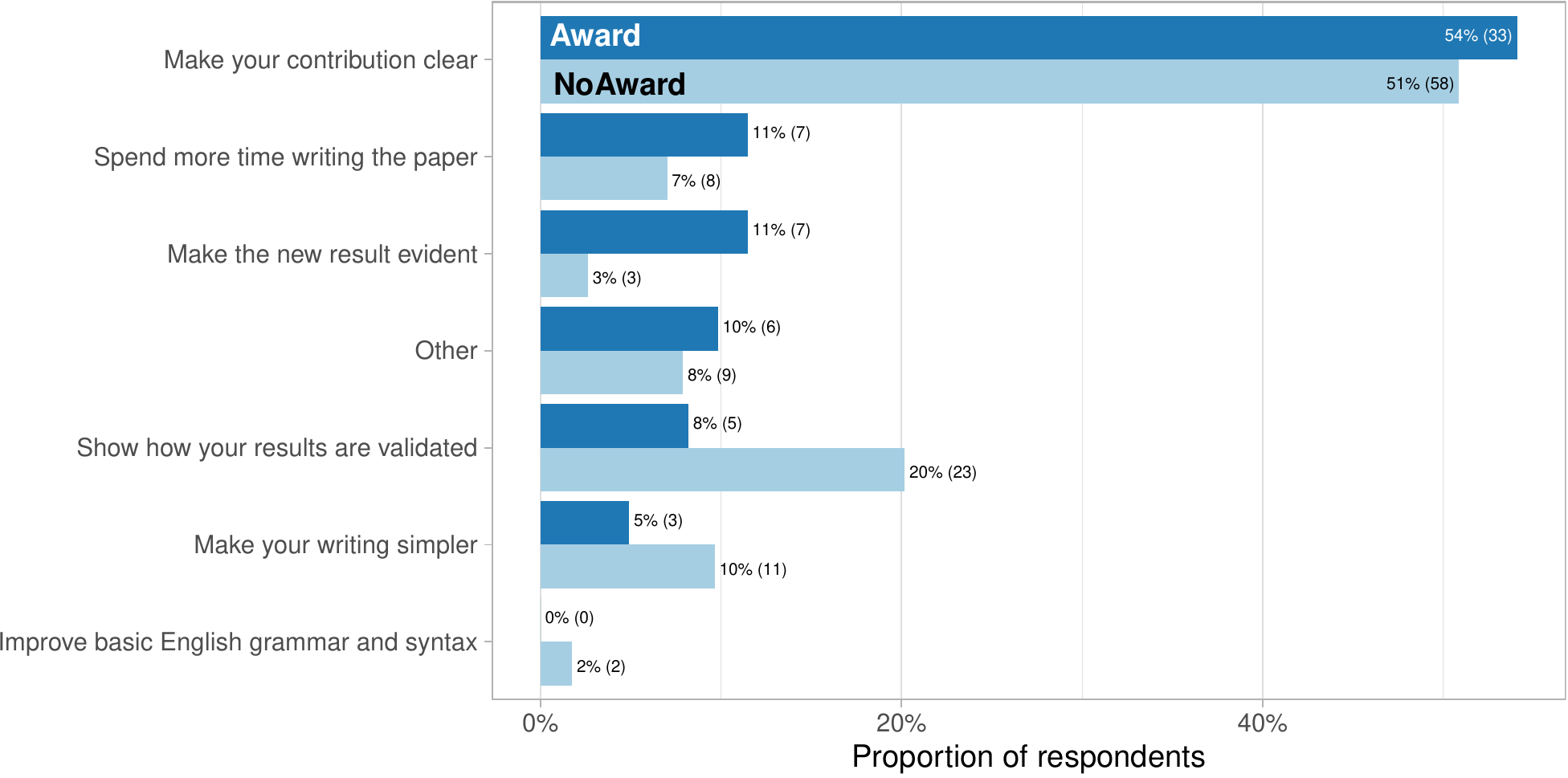}
    \caption{Recommendations by reviewers to increase the chance of acceptance.}
    \label{fig:increase-acceptance}
\end{figure}

Finally, we asked the respondents what recommendations they would give to authors to increase the chance of acceptance.
As Figure~\ref{fig:increase-acceptance} shows, the picture is very clear with little difference between having received an award or not. Authors need to make their contribution clear, which is the recommendation of over half of our respondents.

\paragraph{Discussion---Paper Characteristics} 
Figure~\ref{fig:Char-negative-reviews} and Figure~\ref{fig:Char-positive-reviews} together describe the factors that produce favorable (or, ``not unfavourable'') reviews and answer the research question. These factors align well with common software engineering conference guidelines, and scientific papers more generally: the maxim ``is it new, and is it true''~\cite{macauley12} is confirmed. As an engineering discipline with a heavy focus on tools, a clear and supported validation is highly ranked. However, it is also good see the importance of industry relevance and reproducibility and supporting material. Happily, few reviewers suggested ``citations to my work" as an important characteristic! 

There is a tension between clarity/soundness and interest/novelty. 
The open ended comments and interviews showed that reviewers find perfectly clear and well-executed papers can also be less interesting. Award winning reviewers, in particular, seem to emphasize novelty more.
For some in the software engineering community, their attitude might be summarized by our survey respondent: ``we should encourage the new ideas or problems, even with not enough evidence." Others, however, find more in common with this quote: ``SE is replete with non-controlled `experiments', unvalidated scales, convenience sampling, superficial qualitative research and method slurring".

\subsection{Other Results and Emerging Paradigms}
\paragraph{Replication and reproducibility} % Q20
One question we did not discuss in detail is the role of replication packages. Replication packages are collections of code and data that aid others in reproducing the results of the paper~\cite{fernndez2019open, Shepperd2018}. 
%The claim for replication is that replication packages will strengthen the claims of a paper. 
Replication in software engineering is particularly challenging both for authors to produce and for reviewers to evaluate~\cite{fernndez2019open}.
In the context of peer review, a replication package is another artifact that is subject to review. 
In an engineering/design science discipline like software engineering, replication is a further validation of the technical solutions. However, as one interviewee comments, ``the current way papers are written makes it difficult to reproduce and thus validate the artefacts$^\mathsf{P4}$".
% In cases where the publication venue requires a replication package, then reviewers must consider it. 
The extent to which replication packages are reviewed is unclear. 
In artifact evaluation tracks, which host volunteered replication artifacts from accepted papers, review is thorough and consists of reproducing the results and examining the code and data in depth. 
In other tracks, it is subject to current debate whether we can expect the reviewers to also evaluate to what extent replication is possible.
Table~\ref{tbl:interview-repl} summarizes the interviewee responses about their approach to replication. 
In general, it is uncommon to actually perform a replication, unless a very simple script exists to support it. 

\begin{table*}
    \centering
        \caption{Interviewee replication approaches} 
        \label{tbl:interview-repl}
    \rowcolors{2}{gray!25}{white}
    \begin{tabular}{cp{0.85\linewidth}}
        \toprule
%        \rowcolor{gray!50}
        \textbf{Code} & \textbf{Replication Approach}  \\
    \midrule
    P1 & Rare to run code. Check data against claims in paper for sanity check on methods chosen. \\
	P2 & Check match between raw data and paper claims. Re-run results if script provided.  \\
	P3 & Try opening the data (when available) to clarify possible doubts. \\
	P4 & Brief sanity check for flaws  \\
	P5 & Check when data is interesting, or for critical claims or open questions  \\
	P6 & Is it online? Difficult to use? Don't actually run it \\
	P7 & Check data and scripts available. Do not re-run. \\
	P8 & Check data is available, check statistical maturity  \\
        \bottomrule
\end{tabular}
\end{table*}   

The results of the survey were similar. The majority of respondents, independent of their award status, check for data availability and consistency with the paper (e.g., that the reported sample size matches). 
Only a quarter of reviewers seem to be running code on their machine. A few respondents commented on the difficulty that a double-blind reviewing model poses for this type of evaluation. 
However, there are mechanisms to make double-blind and artifacts work together.\footnote{A good tutorial on data disclosure when using a double-blind review process is provided by Daniel Graziotin: 
\url{https://tinyurl.com/DBDisclose}.
%\url{https://ineed.coffee/5205/how-to-disclose-data-for-double-blind-review-and-make-it-archived-open-data-upon-acceptance/}
} 

\paragraph{Reviewing paradigms}

%%\marco{In my view, the first three paragraph below do not belong here, they should go to the discussion section. An probably a little bit should be anticipated in the intro...}

%Q28
Concerning the review paradigms, we asked our participants about their preferred reviewing model, given existing evidence that single-blind reviewing has implicit biases~\cite{Seeber_2017}. Most respondents (41\%) preferred double-blind, or its extreme version, triple-blind (17\%), with only 27\% favoring single-blind, and just 14\% fully open reviews.

%\begin{figure}[hbt]
%\centering
%  \includegraphics[width=.7\textwidth]{figures/review model overall-1.pdf}
%  \caption{Preferences for review models}
%  \label{fig:reviewmodel}
%\end{figure}

%Discussion
There has long been concern over the direction and state of peer review in software engineering research (and many other research areas). However, with the use of self-archiving (``preprint") servers becoming more and more the norm, the idea of open reviews (where reviewer identity is known), and post-publication peer review are emerging trends to keep in mind. 
%In this subsection, we touch on each of these topics briefly. However, because we did not gather data to provide detailed insights into each of these topics, we leave a more complete analysis for future work.

Preprint servers like arXiv allow for rapid dissemination of results, well in advance of what can be a much lengthier process in journal publication. However, these papers receive only cursory inspection before being deposited on the server, and should not be relied on until reviewed (although peer review does not guarantee quality). % and in several cases have caused major policy changes despite being rife with mistakes. 

Open reviews~\cite{Wolfram_2020} are an emerging open science approach that promises to improve the discourse between author and reviewer. This is due to non-anonymity holding reviewers more accountable for their statements and tone and since both parties know each other's identity. We argue that this openness may lead to more substantial and constructive reviews, at the expense of potential power imbalances between reviewer and author. %There are challenges with power imbalance but we already observe some reviewers who sign all their reviews, blind or otherwise.

%% MARCO: this figure already appears before, it is Fig. 10 in section 3.3 - RQ2
%
%\begin{figure}[hbt]
%  \includegraphics[width=.8\textwidth]{"figures/tool validation-1"}
%  \caption{Approach to replication packages}
%  \label{fig:replication}
%\end{figure}
%

\section{Proto-Guidelines For Reviewing in Software Engineering Venues}
\begin{figure}[htb]
    \centering
    \includegraphics[width=\linewidth]{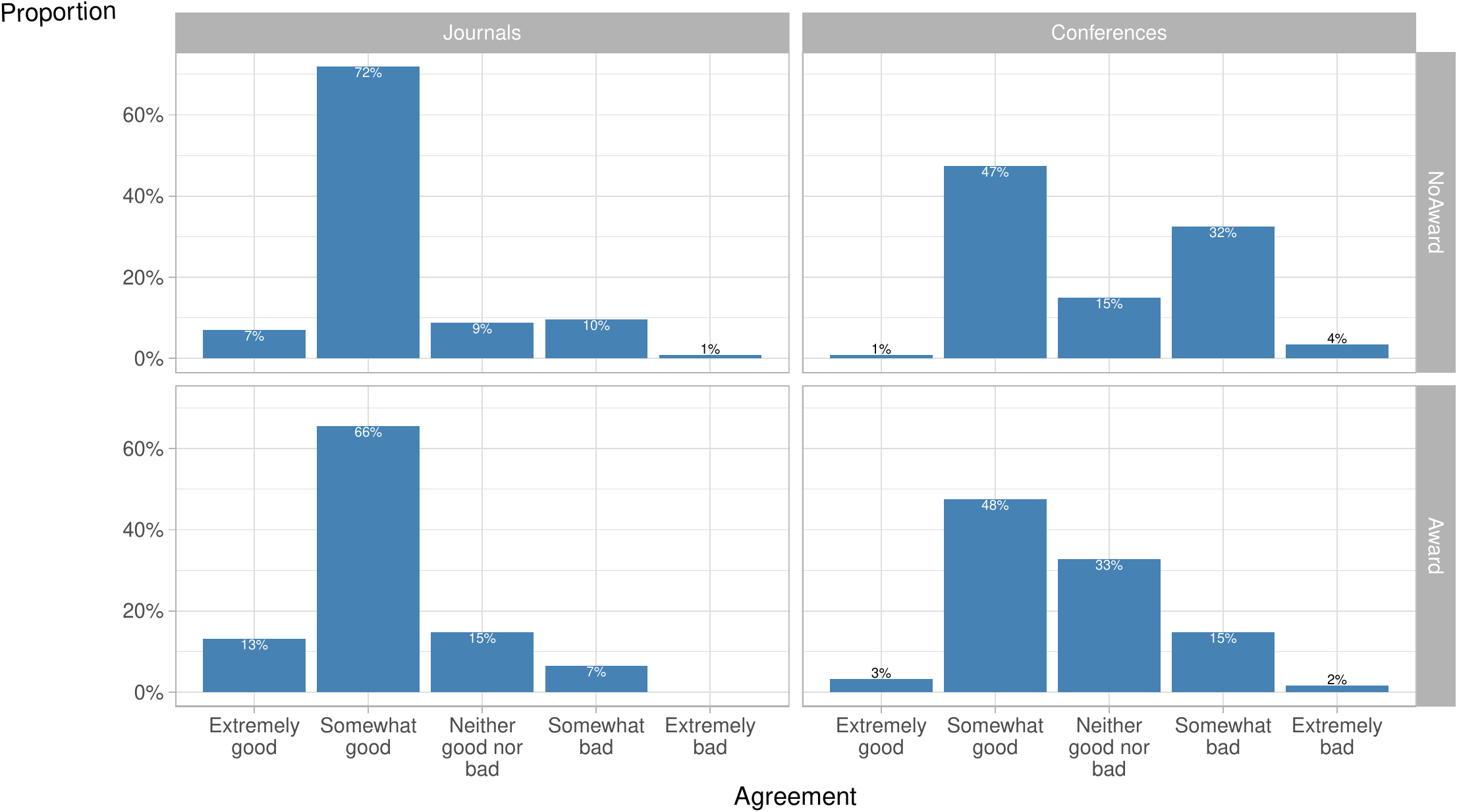}
    \caption{Perceived review quality, per venue/awardee. Red line reflects mean rating.}
    \label{fig:perceive-quality}
\end{figure}

%make such a discussion more actionable, giving, on the basis of what they have learned from data, a set of insights helping reviewers to behave better (it could be a list of patterns/antipatterns, suggestions for reviewing with different expertise, and where possible suggestions about improving the process).
%Provide clear and actionable insights e.g  (proto) guidelines, often directed/separated for different target groups (say researchers/authors, reviewers, editors/journals in this case, as examples).
We asked reviewers about their perception of the level of quality of peer review in software engineering in late 2018 (the date of the survey). Figure~\ref{fig:perceive-quality} shows the respondents believe the state was ``somewhat good''. Award-winners are slightly more optimistic, but this result is likely an artifact of the random sample. In both cases, respondents are more appreciative of reviews received in journals, rather than conferences.

We list some patterns from the data that seem most effective in the process of peer reviewing. These patterns are not research method-specific reviewing guidelines, as these are already under development by SIGSOFT. The SIGSOFT reviewing guidelines~\cite{Ralph2020} have reviewing criteria for different research approaches such as for controlled experiments, e.g., the essential criterion that the study ``justifies how the dependent variable is measured". This way of framing reviewing guidelines in a pattern-based manner can be also found in other initiatives such as artifact evaluation tracks of conferences or open science initiatives in journals.\footnote{See, e.g., artifact evaluation track of ICSE 2021 \url{https://doi.org/10.6084/m9.figshare.14123639} or the open science initiative of the EMSE journal \url{https://github.com/emsejournal/openscience}}

We frame our suggestions according to the template provided in the SIGSOFT empirical standards.\footnote{\url{https://github.com/acmsigsoft/EmpiricalStandards}} We expand on, and provide evidence for, the ``general standard'' that appears at the beginning of that document. The template includes sections for desirable and mandatory criteria for the item the standard applies to, antipatterns, invalid critiques of a paper/process, and general notes and references. We focus on
\bi
    \item the process of reviewing at journals and conferences, i.e., paper assignment and bidding.
    \item the process of reviewing a paper, i.e., reading it and writing a review.
\ei

While we do not provide \emph{writing} guidelines (plenty of excellent advice exists, such as such as Scimel's ``Writing Science"~\cite{schimel}, or Shaw's ``Writing Good SE Research Papers'' \cite{Shaw:2003}), authors who consider these results when writing a paper should increase their chances for paper acceptance.

\subsection{Guideline for the Process of Getting and Managing Reviews}
\subsubsection*{Application}
This guideline is focused on mechanisms for getting reviews and deciding on paper acceptance. It can be used when program chairs or editors adjudicate submitted papers for inclusion in a program or journal issue.
\subsubsection*{Specific Attributes}
Essential Criteria:
\begin{itemize}
	\item  \emph{Insist on reviewer expertise}. Reviewers sometimes bid on papers based on interest, but expertise is more useful for decision making.
	\item \emph{Manage reviewer load}. Review workloads are increasing and seem to be distributed unequally, possibly log-normally. Ozkaya's back-of-napkin analysis \cite{Ozkaya_2021} suggests 30,000 reviews are needed per year in the SE community. If reviewers spend two hours or more merely doing the review, a top-tier conference PC might involve at least 10*2 = 20 hours of work, in addition to the discussions.
	\item The \emph{content of the review is more important} than the conduct and attitude of the reviewers themselves. 
\end{itemize}
Desirable Criteria:
\begin{itemize}
	\item Use a doubly anonymized reviewing model as it is now widely accepted and reduces bias~\cite{Tomkins2017}.
	\item Reviewer load should reflect reviewer submissions. Select reviewers from previous submitters who have not reviewed in the past, or use a self-nomination form to broaden the pool.
	\item Support artifact and replication package initiatives.  Open science initiatives like these are increasingly important.  Will every paper have to justify the lack of a replication package? Will every replication package have to undergo an additional review and if so can we expect the same reviewers who review the papers to also review the disclosed data, material, and source code? We also refer readers to the proposed standard on artifact tracks.\footnote{\url{https://github.com/researchart/patterns/blob/master/standards/artifact.md}}
\end{itemize}
\vspace{2mm}
\begin{flushleft}Extraordinary Criteria:\end{flushleft} 
It is not clear yet what extraordinary chairs and editors are doing. More study is needed here, but from the author perspective,  actionability of the review, and fairness are of primary importance. 

\subsubsection*{General Quality Criteria}
One of the jobs of the editors or program chairs is to ensure reviews are useful to the authors, since in addition to creating the program or issue, the decision-maker should be concerned with the development of the research community (and future submissions to the respective events). Good reviews offer constructive criticism, are justified and factual, and clear. Reviews that do not offer these are helpful for neither the chair/editor nor the authors.
\subsubsection*{Antipatterns}
We do not yet have good data on what anti-patterns look like. One anti-pattern may be failing to accommodate changing topics, or allowing sub-topics and cliques to dominate, since SE topics change considerably over time~\cite{Mathew_2019}. 
% cliques and in-crowds, topic dominance (Tim's paper)
\subsubsection*{Acceptable Deviations}
If sub-reviewers (typically PhD students) are involved in the review process, the reviewer should also review and then meet with the student to synchronize a single review.
\subsubsection*{Invalid Criticisms}
While there is a lot of concern with collegial language, reviews that are harsh yet justified should still be allowed. Cultural differences often factor into how reviewers operate and the chair/editor should ensure these are accommodated.
\subsubsection*{Notes}
Reviews from journals are more highly regarded by authors. Venues should consider why this is the case (typically, the fact there are multiple cycles plays a vital role) and possibly add rebuttals or revision cycles. Turn-around time is important to authors, one reason why feedback on preprints is increasingly important.

Enthusiasm for a paper and its novelty (in context of related work) were highly ranked as leading to more positive reviews. We speculate part of the reason for this perception is because reviewers encounter so many papers a year. The bidding process should identify this potential issue and emphasize the need for expertise in the topic, rather than interest.

Reviewers are weary of incrementalism, e.g., ``Publish less but more significant contributions" and suggested internal review and reflection before submission, when the paper is done, not because the deadline is approaching. 

Random chance plays a role in the process \cite{nips14}. The papers that are neither clear accepts nor clear rejects might easily get their outcome reversed if the reviews were re-assigned. 

\subsubsection*{Exemplars}
\begin{itemize}
	\item Nierstraz's ``Identify the Champion" \cite{Nierstrasz:1998aa}
	\item ICSE final reports from the PC Chairs\footnote{\url{http://www.icse-conferences.org/reports.html}}
	\item NeurIPS 2020 reflections\footnote{\url{https://neuripsconf.medium.com/what-we-learned-from-neurips-2020-reviewing-process-e24549eea38f}}
	\item Journal editorials, such as Jeff Offut's\footnote{\url{https://cs.gmu.edu/~offutt/stvr/17-3-sept2007.html}}, or Ipek Ozkaya \cite{Ozkaya_2021}
\end{itemize}

\subsection{Guidelines for Writing Reviews}
\label{sec:writingguide}
\subsubsection*{Application}
This guideline is for general characteristics of well-written peer reviews in software engineering. Specific review requirements are topic- and method-specific, and dedicated guidelines, such as on sampling approach, use of open coding, or experimental analysis, can be found in the SIGSOFT empirical review standards \cite{Ralph2020}.

\subsubsection*{Specific Attributes}

Essential Criteria:
\begin{itemize}
	\item \emph{Stay constructive and helpful}. It is of no use to write a review that does not provide indicators for improvement. Even for a study that is not salvageable, there is an educational aspect to explaining how the authors could do it properly (e.g., follow community norms, improve clarity, improve the study). %The authors have typically expended enormous effort to contribute to our community.
	\item Statements included in a review should be \emph{factual and justified}. Examples of this are to point out the precise figure or claim being criticized and to provide specific references when pointing out missing related work.
	\item A reviewer should be \emph{fair} and adequately balance negative and positive aspects. 
	\item Identify clearly the \emph{reason of criticism}, i.e., if support for claims was missing, or if the method used was poorly matched with the knowledge claim.
\end{itemize}
Desirable Criteria:
\begin{itemize}
	\item Consider whether language matters as much as you think. Not every writer is a native speaker, and typos and grammatical infelicities are unlikely to detract from the main ideas.
	\item The attitudes of a good reviewer should be \emph{accuracy and fairness}. Reviewers should read the paper carefully, entirely, and not in a rush.
	\item Techniques such as PDF annotation on a tablet were frequently used by more expert reviewers and may facilitate the review process.
	\item Artifacts that form an essential part of the paper's claims should be examined at least superficially, i.e., for availability, consistency, and whether it is well documented. One recommendation is to refer to existing reviewing guidelines for artifact evaluations and, at least, briefly compare the submission to the criteria listed therein. 
	\item Allocate more than 2 hours for reviewing a journal paper, and from 1-3 hours for a conference paper, including the writeup. Most reviewers tend to read the paper once entirely, with some needing 2-3 readings.
\end{itemize}

\begin{flushleft}Extraordinary Attributes:\end{flushleft} %what the award winners do based on R script
\begin{itemize}
	\item Award winning peer reviews tend to be detailed and actionable. 
	\item Excellent reviews are tightly focused on the venue guidelines. They go into detail on the associated artifacts, checking them for consistency and availability. 
	\item Award winners focus more on novelty and soundness over practical relevance. 
%	\item Excellent reviews are clear, factual, and accurate.
\end{itemize}

%\paragraph{General Quality Criteria}
\subsubsection*{Antipatterns}
Poor reviews are rushed; overly personal (e.g., directly addressing authors, rather than their work, often suggesting ad-hominem rejections); and tend to be short, or, if longer, focused on syntax and editorial objections. Poor reviews do not point out related work in detail (e.g., with a DOI), and fail to consider the scope of the paper relative to the review guidelines. 

While novelty and interest are naturally appealing, the literature suggests reviewers are poor at identifying truly novel results~\cite{Horbach_2018} or able to judge what might be of interest to the wider community. 
\subsubsection*{Acceptable Deviations}
 Although many guidelines from chairs emphasize kindness, this is less of a concern than the preceding attributes. Rather than kindness, collegiality and helpfulness are more important.
 Some criteria may be relaxed when reviewers are conducting a rapid review (e.g. when serving as an additional reviewer to add to ongoing discussions) or when evaluating a paper in a unique venue, e.g., for a doctoral symposium or tools paper.
\subsubsection*{Invalid Criticisms} % of a review
Peer review is not good at identifying all possible errors in a paper. There are plenty of flawed, yet peer-reviewed papers, so a good review should not attempt, nor be expected to, find every single error. 
\subsubsection*{Notes}
A good peer review helps the journal editor or program chair as much as the author of the paper. Individual venues may have different criteria the reviewer should follow beyond these guidelines.
It is reasonable to assume that the authors have put an enormous amount of time into their research and in writing the paper. Do not assume authors are ignorant or lazy.

\subsubsection*{Exemplars}
As peer review is typically a confidential process, we can only point readers to existing open reviews. The reviews on the PeerJ paper on gender differences in pull request acceptance \cite{Terrell_2017} are good examples.\footnote{\url{https://peerj.com/articles/cs-111/reviews/}}. Another resource is the reviews on OpenReview.net, e.g. \url{https://openreview.net/forum?id=rklXaoAcFX&noteId=HyeF-4_9hm}

%\section{Discussion}
%First, we look at how reviewers understand replication issues in papers. Then we consider new reviewing paradigms. Finally, we consider our study limitations. 
%

\section{Threats to Validity}
\label{sec:discussion}

\textit{Construct validity}: concerns the measurement of the constructs in a proper way.
%	Who makes the list of “great reviewers”?  \\
%How were venues/list selected? \\
The people we chose for the initial interviews introduces bias. Nevertheless we do not consider this bias a threat for three main reasons: (i) the four of us---though all working in the area of software engineering---have different backgrounds and research topics; (ii) the interviewees provided a consistent view on reviewing with clearly different points of view; (iii) we used the outcomes of the interviews as a starting point---mediated by the authors---to build the general conceptual framework underpinning the questionnaire instrument.  

Question wording might result in subjective responses, in particular if the respondent understood the questions differently than we had intended. We validated the questionnaire with pilot studies to mitigate this potential threat. 

\noindent\textit{Internal validity}: concerns the correct definition and conduct of the procedures.
We assume the survey and interview responses are truthful. However, it is possible that a few options have been under- (or over-) reported.
This potential bias is more likely for socially undesirable (or desirable) behavior.
For instance, we had only one respondent who reported that citations to his/her own work may lead to a positive review. This response is a clear example of socially undesirable behavior that might have been under-reported. Nevertheless, that item seems to be a rather isolated example for what we would judge as socially undesirable behaviors. 

Self-reported productivity and effort are potentially biased~\cite{Zerbe1987}. Thus, respondents may inflate their estimates of how long a review takes or how many reviews they perform per year. In the future, we could correct for this potential estimation error by cross-referencing self-reports with journal statistics from editors. However, in general, it is ``a truth universally acknowledged'' in our field that the number of reviews only increases each year, we judged this risk to be minimal, and in any case, not germane to our main discussion.

\noindent\textit{Conclusion validity}: refers to drawing correct decisions from the data.
In our analysis of the survey results, we did not perform any statistical hypothesis tests because we still lack a proper theory on peer reviewing in software engineering.
Because of the nature of this study, we comment on proportions and on the ordering of alternatives. 
Based on the sample size we can estimate a confidence interval for proportions that is roughly $\pm 7\%$. Therefore we avoided considering differences smaller than such threshold.

\noindent\textit{External validity}: affects the possibility to generalize the results.
We gathered the list of participants from a wide range of software engineering conferences and journals that cover a broad range of software engineering topics. Therefore, we believe we obtained a general picture of the software engineering community. With our approach it is possible we overlooked peculiarities specific to sub-communities.

The response rate is 23\%. However, considering only the responses we included in the analysis, that figure drops to 18\%. We acknowledge this response rate might limit how representative the sample is of the larger software engineering community. However, such response rates are consistent with most surveys conducted in our discipline. Our sub-group numbers are quite small (e.g., for those with less than 5 years experience), so we do not advise generalizing to the wider population. For instance, we might have missed poor reviewers, if one of the reasons for poor reviews was lack of time, which might also affect their response to a survey request.
%now just limitations
\section{Related Work}
\label{sec:related}

Peer review has long been of interest to the scientific community. Spier's paper~\cite{Spier2002} on the history of peer review is a fascinating read on its emergence. More recently, Tennant et al.~\cite{Tennant2017} have examined the more recent history and future directions in peer review. Emerging trends include the degree of anonymity, the way in which to reward peer reviewers, and how to reduce bias. In the remainder of this section, we list a few studies and reports directly related to the field of software engineering. Those results are in line with a study conducted by Prechelt et al.~\cite{PrecheltGF18}, surveying the ICSE community on the future of peer review (the open data set is available\footnote{\url{https://doi.org/10.6084/m9.figshare.5086357.v1}}).
%https://figshare.com/articles/Open_Science_Repository_for_On_the_Status_and_Future_of_Peer_Review_in_Software_Engineering_/5086357}}).

The broader area of peer review is the focus of scientometrics research~\cite{Squazzoni_2017}, which has looked at several relevant areas: whether open peer review improves paper quality~\cite{Zong_2020}, whether peer review is effective at catching flawed papers (not always~\cite{Horbach_2018}), and the effectiveness of single- vs double-blind reviews~\cite{Seeber_2017}. Ragone et al.~\cite{Ragone_2013} looked specifically at computer science venues, conducting a data analysis of reviews and paper scores in 10 different CS conferences. They concluded that review efficiency (a metric of review time and paper length) could be optimized by identifying earlier which papers are likely to be rejected (e.g., after 2 reviews). Our paper focuses on reviewer practices and contributes qualitative analysis of how reviewers themselves perceive the process.

An early paper by Alan Jay Smith, ``The Task of the Referee"~\cite{Smith:1990aa}, lays out reviewer guidelines. Other editors and chairs often encourage better peer review, for example Rubin~\cite{drubin11}. Nierstrasz's article ``Identify The Champion"~\cite{Nierstrasz:1998aa} contains detailed guidance on how to structure program committee discussions, and move from individual reviews to program decisions. The \emph{champion} is a reviewer who is motivated to see the paper succeed.

The alternate approach to guidelines for reviewers is to provide guidelines for authors on how to navigate reviewers. Pre-eminent in software research is the work of Mary Shaw~\cite{Shaw:2003}, and the subsequent follow-up by Theisen et al.~\cite{Theisen2017}. As for peer review practices from the empirical point of view, the study by Prechelt et al.~\cite{PrecheltGF18} looked at perspectives on the current state and future of peer review in Software Engineering. They report on the author perspective on peer reviews. We build on their work by directly reporting on the reviewer perspectives on peer review, and what a good review looks like.

A recent paper (2020) by Petre et al.'s working group on peer review in CS Education~\cite{Petre_2020} sought to ``map the landscape of peer-review practice in computing education research and to seek insights about what influences decisions about the process and criteria.''
Using interviews, surveys, and document analysis, that paper identified criteria, roles, and etiquette expected of (and in) peer review. For example, they present the following important paper criteria (our explanations in square brackets):
\begin{itemize}
	\item Author Ethics [plagiarism, honest about limitations]
	\item Effective Presentation 
	\item Real-World Impact
	\item Novelty/Contribution
	\item Sufficient Detail 
	\item Rigor/Sound Methodology
	\item Situation in Prior Work 
	\item Important Question 
	\item Scope [applicable to venue]
\end{itemize}

Our paper has focused on a different community (although with some overlap), and looked in more detail at the specific practices and criteria individual reviewers applied to papers. The study by Petre et al., however, extends this work by examining the process by which \emph{venues} look at reviewing, something we did not examine in detail.

%The NIPS experiment in 2014 was an attempt to understand consistency in peer reviewing. Using parallel program committees reviewing the same papers, about 57\% of the papers accepted by the first committee were rejected by the second one and vice versa \cite{nips14}.% They found that from 38\%-64\% of the actual presented papers would be re-selected in a second, identical run of the conference, with a new program committee. 
%A Jupyter notebook explaining the reasoning and data is available.\footnote{\url{https://github.com/sods/conference/blob/master/The\%20NIPS\%20Experiment.ipynb}}

\subsection{Experience reports and Community Initiatives}
The grey literature---blogs, commentaries in journals, editor's notes, and so on---are, in our view, still the best current source for explicit guidelines on how to do peer review. 
Several eminent scientists have reported on their best practices. As an example, we point out Alexander Serebrenik, from our community, winner of multiple best reviewer awards, who summarized his approach in a set of slides\footnote{\url{https://www.slideshare.net/aserebrenik/peer-reviews-119010210}}. We list other references in our replication package.

There are initiatives to improve the quality of peer review. Lutz Prechelt's \textit{Review Quality Collector} tool\footnote{\url{https://reviewqualitycollector.org}} is a measure that integrates with the EasyChair platform to enable tool-assisted assessment of peer review quality. One of us (NE) used this when acting as PC chair to collect information on who the best reviewers were. The tool supports adjustable criteria and criteria weights, including timeliness, helpfulness to authors, and helpfulness to PC chairs. The associated website has a long list of articles on peer review and its trials and tribulations\footnote{\url{http://www.inf.fu-berlin.de/w/SE/ReviewQualityCollectorHome}}.

Venues across computer science disciplines experiment with peer review on a fairly regular basis: the above-mentioned NeurIPS experiment; double blind reviewing at SIGMOD, a main venue for database research~\cite{Tung_2006}; merging conferences and journals at VLDB and PVLDB~\cite{andVanessaBraganholo:2016aa}; cyclical reviews at the conference on Computer-Supported Cooperative Work~(CSCW)\footnote{http://cscw.acm.org/2019/CSCW-2020-changes.html}; and an analysis of peer review in CS education~\cite{Petre_2020}, to name a few.

ACM SIGSOFT's peer review initiative~\cite{Ralph2020}, edited by Paul Ralph, is an initial collection of guidelines for reviewers of common research methods, such as survey research or controlled experiments.  The goal is to ensure reviewers understand the current best practices for a particular research method. Having a set of guidelines for reviewing different research methods is complementary to our focus on what makes a good review, since in our results, reviewer expertise in the topic and method was seen as vital. 

A related effort aims to improve the review process itself, such as with open reviews and retrospectives. Open reviews (e.g., OpenReviews.net) are post-publication reviews in which the reviewer's identity is usually known to the author.
The International Conference on Software Engineering conducts post-mortems of each year's review process. Although primarily focusing on the mechanism for deciding on paper acceptance, these reports also explain types of papers accepted and what review information was important in making a decision\footnote{\url{http://www.icse-conferences.org/reports.html}}. A recent series by Jacopo Soldani and colleagues in ACM SIGSOFT's Software Engineering Notes~\cite{Soldani_2020} summarizes the ``pains and gains" of peer review and provides discussion about the concerns many scientists have with peer review, such as workload.

\subsection{Peer Review Criteria}
It is common for a peer review request to include some criteria by which the paper is to be judged. These criteria are particularly important when the paper is a non-standard (i.e., non-technical research) format. For example, the \textit{Empirical Software Engineering Journal} allows for Technical Papers and Engineering in Practice papers. The \textit{Journal of Systems and Software} accepts technical papers, New Ideas and Trends, and In Practice papers. The \textit{International Conference on Software Engineering (ICSE)} has numerous tracks, including Software in Society, Software Engineering in Practice, the main technical track, New Ideas and Emerging Results, Software Engineering Education and Training, and more. 

Most conferences have the same or similar reviewing criteria for their research tracks, typically summarized along the call for papers. The ICSE technical review criteria, for instance, are available on the conference call for papers. For the 2021 edition, the criteria\footnote{\url{https://conf.researchr.org/track/icse-2021/icse-2021-papers\#Call-for-Papers}} are 
\begin{itemize}
	\item Soundness: The extent to which the paper’s contributions are supported by rigorous application of appropriate research methods.
	\item Significance: The extent to which the paper’s contributions are important with respect to open software engineering challenges.
	\item Novelty: The extent to which the contribution is sufficiently original and is clearly explained with respect to the state-of-the-art.
	\item Verifiability: The extent to which the paper includes sufficient information to support independent verification or replication of the paper’s claimed contributions.
	\item Presentation: The extent to which the paper’s quality of writing meets the high standards of ICSE [...].
\end{itemize}

%As an example of reviewing criteria for new ideas tracks, here is one extracted set, from ICSME 2021:
%\quote{Please take special care to treat the papers as what the are: new ideas and emerging results. A good evaluation is *not* a factor for accepting the paper. These are not just short papers that did not make it to the research track. This is the track that gives space to cool and crazy or insightful ideas that can move the field of software maintenance and evolution forward. Yet, they are not (or only partially) implemented and usually not empirically evaluated. Hence, in contrast to a normal research track, novelty is most important. So our main criteria for evaluation are:
%Relevance for the conference topics, Novelty, Quality of presentation/discussion of the idea. Such tracks commonly emphasize the topic and its novelty over rigorous validation.

%Besides submitting your reviews on time, we appreciate your participation in the online discussion as soon as more than one review is available.
%}

Common to most review guidelines is a focus on presentation and clarity of the paper, adherence to the conference or journal topics, and length. Increasingly guidelines are also emphasizing the importance of constructive criticism (the Reviewer 2 problem~\cite{Peterson2020}) and trying to see what is positive and worthwhile about the paper (``please look for reasons to accept a paper and to praise authors for what they did well"). However, we did not find this concern highly ranked in our dataset.

\section{Conclusion}
%\subsection{Summary of conclusions}

%% MARCO: moved to discussion
%\subsection{Threats to Validity}
%Usual survey/interview limitations
%
%	Who makes the list of “great reviewers”?  \\
%How were venues/list selected? \\
%Difference between reviewers, editors/PC chairs, authors, students vs senior faculty.
%Some questions limited responses.

%% REMOVED consideration about open servers to discussion about emerging paradigms

This study conducted an interview and survey-based study to understand some of the mechanisms by which reviewers in software engineering conduct reviews. We secured a large cohort of respondents derived from program committee members and journal reviewers of the past few years, many of whom have received recognition for their excellent reviews. One respondent summarizes our motivation: ``[It] seems very straightforward to do a good review, but given my [experience managing reviews] I see that it is not happening very often$^\mathsf{P6}$.

Our paper outlined what these respondents see happening in the peer review world, including a large amount of time spent on reviews.  We then explained the practices common to these respondents, followed by some of their tips and beliefs about what should be in a review, and how paper writers can optimize their paper for reviewers. Most important properties of a good review are for it to be helpful and factually-based. While we initially thought the perceptions and practices of the award-winners would differ from non-award winners, the results show remarkably few large differences. 

Our results are encouraging. Our respondents were generally happy to perform their work, and saw it as an essential (albeit uncompensated) part of their scientific career. However, problems exist with review workload and subject-matter expertise, which directly impact the ability of reviewers to continue high-quality review work.

Finally, we concluded our manuscript by compiling a guideline for reviewing, reflecting the insights we gained through our study. This guideline is naturally far from being perfect, let alone complete. Yet, our hope is to make one additional contribution to the ongoing effort to further improve our peer review models in the future.

%Finally, since the majority of research receives little attention (in the form of citations), one might ask if pre-publication review is necessary for all papers. After all if no one relies on the result, its quality is ultimately a moot point. Post publication review would thus focus on the papers that interest reviewers the most. This type of review already occurs, of course, as much-heralded papers received attention on social media. However, the problem remains identifying the quality among the noise, for which pre publication review seems to have few alternatives.

%\newpage
\printbibliography 

%\section{Appendix - Data analysis}
%neil - move appendix to replication package
%\appendix
%\input{questionnaire.tex}

\end{document}